\def\d0{D\O}
\def\dedge{$d_{\rm edge}$}
\def\epem{$e^+e^-$}
\def\edgec{$\widetilde{\rm C}$}
\def\edgecc{$\widetilde{\rm C}$C}
\def\edgeboth{$\widetilde{\rm C}\widetilde{\rm C}$}
\def\edgece{$\widetilde{\rm C}$E}
\def\fedge{$\widetilde f$}
\def\alphaedge{$\widetilde \alpha$}
\def\deltaedge{$\widetilde \delta$}
\def\sedge{$\widetilde s$}
\def\cedge{$\widetilde c$}
\def\nedge{$\widetilde n$}

\def\ptw{$p_T(W)$}
\def\pte{$p_T(e)$}
\def\ptnu{$p_T(\nu)$}
\def\mt{$m_T$}

\def\upar{$u_\parallel$}
\def\met{\mbox{$\rlap{\kern0.15em/}E_T$}} 

\documentclass[prd,twocolumn,floatfix]{revtex4}
\usepackage{epsfig}

\begin{document}



\title{Improved $W$ boson mass measurement with the D\O\ detector}

%
\author{                                                                      
V.M.~Abazov,$^{23}$                                                           
B.~Abbott,$^{57}$                                                             
A.~Abdesselam,$^{11}$                                                         
M.~Abolins,$^{50}$                                                            
V.~Abramov,$^{26}$                                                            
B.S.~Acharya,$^{17}$                                                          
D.L.~Adams,$^{55}$                                                            
M.~Adams,$^{37}$                                                              
S.N.~Ahmed,$^{21}$                                                            
G.D.~Alexeev,$^{23}$
% ...
\\                                                                            
\vskip 0.30cm                                                                 
\centerline{(D\O\ Collaboration)}                                             
\vskip 0.30cm                                                                 
}                                                                             
\address{                                                                     
\centerline{$^{1}$Universidad de Buenos Aires, Buenos Aires, Argentina}       
\centerline{$^{2}$LAFEX, Centro Brasileiro de Pesquisas F{\'\i}sicas,         
                  Rio de Janeiro, Brazil}                                     
\centerline{$^{3}$Universidade do Estado do Rio de Janeiro,                   
                  Rio de Janeiro, Brazil}                                     
\centerline{$^{4}$Institute of High Energy Physics, Beijing,                  
                  People's Republic of China}                                 
\centerline{$^{5}$Universidad de los Andes, Bogot\'{a}, Colombia}             
\centerline{$^{6}$Charles University, Center for Particle Physics,            
                  Prague, Czech Republic}                                     
\centerline{$^{7}$Institute of Physics, Academy of Sciences, Center           
                  for Particle Physics, Prague, Czech Republic}               
\centerline{$^{8}$Universidad San Francisco de Quito, Quito, Ecuador}         
\centerline{$^{9}$Institut des Sciences Nucl\'eaires, IN2P3-CNRS,             
                  Universite de Grenoble 1, Grenoble, France}                 
\centerline{$^{10}$CPPM, IN2P3-CNRS, Universit\'e de la M\'editerran\'ee,     
                  Marseille, France}                                          
\centerline{$^{11}$Laboratoire de l'Acc\'el\'erateur Lin\'eaire,              
                  IN2P3-CNRS, Orsay, France}                                  
\centerline{$^{12}$LPNHE, Universit\'es Paris VI and VII, IN2P3-CNRS,         
                  Paris, France}                                              
\centerline{$^{13}$DAPNIA/Service de Physique des Particules, CEA, Saclay,    
                  France}                                                     
\centerline{$^{14}$Universit{\"a}t Mainz, Institut f{\"u}r Physik,            
                  Mainz, Germany}                                             
\centerline{$^{15}$Panjab University, Chandigarh, India}                      
\centerline{$^{16}$Delhi University, Delhi, India}                            
\centerline{$^{17}$Tata Institute of Fundamental Research, Mumbai, India}     
\centerline{$^{18}$Seoul National University, Seoul, Korea}                   
\centerline{$^{19}$CINVESTAV, Mexico City, Mexico}                            
\centerline{$^{20}$FOM-Institute NIKHEF and University of                     
                  Amsterdam/NIKHEF, Amsterdam, The Netherlands}               
\centerline{$^{21}$University of Nijmegen/NIKHEF, Nijmegen, The               
                  Netherlands}                                                
\centerline{$^{22}$Institute of Nuclear Physics, Krak\'ow, Poland}            
\centerline{$^{23}$Joint Institute for Nuclear Research, Dubna, Russia}       
\centerline{$^{24}$Institute for Theoretical and Experimental Physics,        
                   Moscow, Russia}                                            
\centerline{$^{25}$Moscow State University, Moscow, Russia}                   
\centerline{$^{26}$Institute for High Energy Physics, Protvino, Russia}       
\centerline{$^{27}$Lancaster University, Lancaster, United Kingdom}           
\centerline{$^{28}$Imperial College, London, United Kingdom}                  
\centerline{$^{29}$University of Arizona, Tucson, Arizona 85721}              
\centerline{$^{30}$Lawrence Berkeley National Laboratory and University of    
                  California, Berkeley, California 94720}                     
\centerline{$^{31}$University of California, Davis, California 95616}         
\centerline{$^{32}$California State University, Fresno, California 93740}     
\centerline{$^{33}$University of California, Irvine, California 92697}        
\centerline{$^{34}$University of California, Riverside, California 92521}     
\centerline{$^{35}$Florida State University, Tallahassee, Florida 32306}      
\centerline{$^{36}$Fermi National Accelerator Laboratory, Batavia,            
                   Illinois 60510}                                            
\centerline{$^{37}$University of Illinois at Chicago, Chicago,                
                   Illinois 60607}                                            
\centerline{$^{38}$Northern Illinois University, DeKalb, Illinois 60115}      
\centerline{$^{39}$Northwestern University, Evanston, Illinois 60208}         
\centerline{$^{40}$Indiana University, Bloomington, Indiana 47405}            
\centerline{$^{41}$University of Notre Dame, Notre Dame, Indiana 46556}       
\centerline{$^{42}$Iowa State University, Ames, Iowa 50011}                   
\centerline{$^{43}$University of Kansas, Lawrence, Kansas 66045}              
\centerline{$^{44}$Kansas State University, Manhattan, Kansas 66506}          
\centerline{$^{45}$Louisiana Tech University, Ruston, Louisiana 71272}        
\centerline{$^{46}$University of Maryland, College Park, Maryland 20742}      
\centerline{$^{47}$Boston University, Boston, Massachusetts 02215}            
\centerline{$^{48}$Northeastern University, Boston, Massachusetts 02115}      
\centerline{$^{49}$University of Michigan, Ann Arbor, Michigan 48109}         
\centerline{$^{50}$Michigan State University, East Lansing, Michigan 48824}   
\centerline{$^{51}$University of Nebraska, Lincoln, Nebraska 68588}           
\centerline{$^{52}$Columbia University, New York, New York 10027}             
\centerline{$^{53}$University of Rochester, Rochester, New York 14627}        
\centerline{$^{54}$State University of New York, Stony Brook,                 
                   New York 11794}                                            
\centerline{$^{55}$Brookhaven National Laboratory, Upton, New York 11973}     
\centerline{$^{56}$Langston University, Langston, Oklahoma 73050}             
\centerline{$^{57}$University of Oklahoma, Norman, Oklahoma 73019}            
\centerline{$^{58}$Brown University, Providence, Rhode Island 02912}          
\centerline{$^{59}$University of Texas, Arlington, Texas 76019}               
\centerline{$^{60}$Texas A\&M University, College Station, Texas 77843}       
\centerline{$^{61}$Rice University, Houston, Texas 77005}                     
\centerline{$^{62}$University of Virginia, Charlottesville, Virginia 22901}   
\centerline{$^{63}$University of Washington, Seattle, Washington 98195}       
}                                                                             

\begin{abstract}
We have measured the $W$ boson mass using the \d0 detector
and a data sample of 82 pb$^{-1}$ from the Fermilab Tevatron collider.
This measurement uses $W\rightarrow e \nu$ decays, where the electron
is close to a boundary of a central 
electromagnetic calorimeter module.  Such
`edge' electrons
have not been used in any previous \d0 analysis, and represent a 14\%
increase in the $W$ boson sample size.   For these electrons, new
response and resolution parameters are determined, and 
revised backgrounds and underlying event energy flow 
measurements are made.   When the current measurement is
combined with previous \d0 $W$ boson mass measurements, we obtain
$M_W = 80.483 \pm 0.084$ GeV.  The 8\% improvement 
from the previous \d0 measurement 
is primarily due to the improved determination of the
response parameters for non-edge electrons using the sample
of $Z$ bosons with non-edge and edge electrons.
\end{abstract}

\maketitle


\section{Introduction}
\label{introduction}

In the past decade, many experimental results have 
improved our understanding of the standard model (SM) \cite{sm}
of electroweak interactions as an
excellent representation of nature at the several hundred GeV scale
\cite{ewwkgp}.
Dozens of measurements have determined the parameters of
the SM, including, indirectly, the mass of the as-yet unseen Higgs boson.
The $W$ boson mass measurement plays a critical role in constraining
the electroweak higher order corrections and thus gives a powerful
constraint on the mechanism for electroweak symmetry breaking.


Recently, direct high precision measurements of $M_W$ have been made by the 
\d0 \cite{ecmw,ccmw1b,ccmw1a}
and CDF \cite{cdfmw}
collaborations at the Fermilab $\overline p p$ collider, 
and by the ALEPH \cite{alephmw}, DELPHI \cite{delphimw}, 
L3 \cite{l3mw} and OPAL \cite{opalmw}
collaborations at the CERN LEP-2 $e^+e^-$ collider.
The combined result of these measurements and preliminary LEP-2
updates \cite{ewwkgp} is $M_W=80.451 \pm 
0.033$ GeV.   The combined indirect determination
of $M_W$ \cite{ewwkgp}
from measurements of $Z$ boson properties at LEP and SLC, taken
together with neutrino scattering studies \cite{nutevmw} 
and the measured top quark mass \cite{topquarkmass}, is
$M_W=80.373\pm0.023$ GeV,
assuming the SM \cite{ewwkgp}.  The reasonable
agreement of direct and indirect measurements is an indication of
the degree of validity of the SM.  Together with other precision
electroweak measurements, the $W$ boson measurement
favors a Higgs boson with mass
below about 200 GeV.
Measurement of $M_W$ with improved precision is of great importance, as
it will enable more stringent tests of the SM, particularly if
confronted with direct measurement of the mass of the Higgs boson,
or could give an indication of physics beyond the standard paradigm.

The  measurements of $M_W$ in the \d0 experiment use
$W$ bosons produced in $\overline p p$ collisions at 1.8 TeV
at the Fermilab Tevatron collider, with subsequent decay $W\rightarrow e\nu$.
The previous measurements are distinguished by the location of the electron
in a central electromagnetic calorimeter ($|\eta_e| \leq 1.1$) 
\cite{ccmw1b,ccmw1a}
or the end calorimeters ($1.5 \leq |\eta_e| \leq 2.5$) \cite{ecmw},
where $\eta$ is the pseudorapidity, $\eta= - \ln \tan \theta /2$, and
$\theta$ is the polar angle.
The measured quantity is the ratio $M_W/M_Z$, which is converted to
the $W$ boson mass using the precision $Z$ boson
mass from LEP \cite{ewwkgp}.  Decays of the $Z$ boson into \epem ~are
crucial for determining many of the detector response parameters.
For all previous \d0 $W$ boson mass measurements (and for other studies of
$W$ and $Z$ boson production and decay), electrons 
in the central electromagnetic calorimeter were excluded if they were 
close to the module boundaries in azimuth ($\phi$).
In this paper, we revisit the central electron $W$ boson 
analysis, adding these hitherto
unused electron candidates that appear near the calorimeter module 
boundaries \cite{slava}.  We use a data sample of 82 pb$^{-1}$ obtained
from the 1994 -- 1995 run of the Fermilab collider.

\section{Experimental method and event selection}
\label{experiment}

\subsection{Detector}

The \d0 detector \cite{d0nim} for the 1992 -- 1995 Fermilab collider run 
consists of a tracking region that
extends to a radius of 75 cm from the beam and contains inner and outer drift
chambers with a transition radiation detector between them.  
Three uranium/liquid-argon calorimeters outside the tracking
detectors are housed in
separate cryostats: a central calorimeter and two end calorimeters.
Each calorimeter has an inner section for detection of electromagnetic
(EM) particles; these consist of twenty-one uranium plates of 3 mm thickness for
the central calorimeter or twenty 4 mm thick 
uranium plates for the end calorimeters.
The interleaved spaces between absorber plates contain signal readout boards
and two 2.3 mm liquid argon gaps.  There are four separate EM readout
sections along the shower development direction.
The transverse segmentation of the EM
calorimeters is 0.1$\times$0.1 in $\Delta\eta\times\Delta\phi$, 
except near the EM shower maximum, where the
segmentation is 0.05$\times$0.05 in $\Delta\eta\times\Delta\phi$.  
Subsequent portions of the calorimeter have thicker uranium or
copper/stainless steel absorber plates and are used to measure hadronic
showers.  
The first hadronic layer is also used to capture any energy
escaping the EM layers for electrons or photons.
The muon detection system outside the calorimeters is
not used in this measurement, except as outlined in Refs. 
\cite{ecmw,ccmw1b,ccmw1a}
for obtaining a muon track sample used to calibrate the drift chamber
alignment.

\begin{figure}[!hbp]
\epsfig{figure=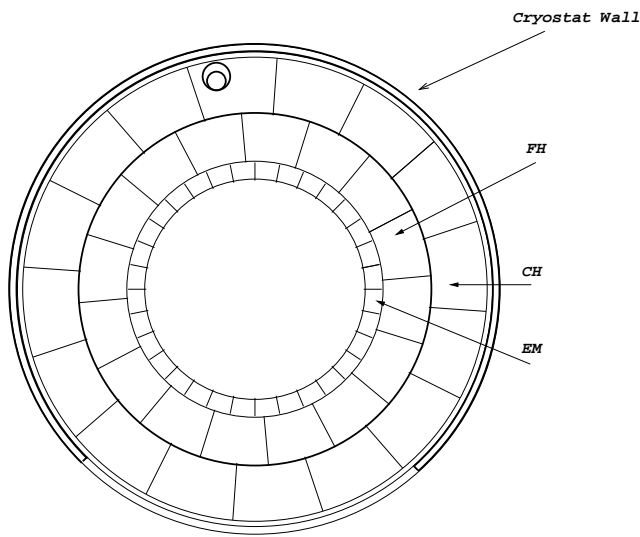,width=8.5cm}
\caption{
End view of the central calorimeter showing the arrangement
for electromagnetic (EM), fine hadronic (FH) and coarse hadronic (CH) modules.
The Tevatron Main Ring passes through the circular hole near the top
of the CH ring.
}
\label{fig:ccendview}
\end{figure}

An end view of the central calorimeter is shown in 
Fig.~\ref{fig:ccendview}.  
There are three concentric barrels of modules;
the innermost consists of thirty-two EM modules, followed by 
sixteen hadronic
modules with 6 mm uranium absorber plates, and then sixteen coarse hadronic
modules with 40 mm copper absorber plates to measure the tails of hadronic
showers.  All previous \d0 $W$ boson mass analyses using central electrons have
imposed cuts on the electron impact position in the EM
modules that define a fiducial region covering the interior 80\% in 
azimuth of
each module.   Such electrons will be referred to in this paper
as `C' or `non-edge' electrons.   The remaining 
central electrons that impact on the
two 10\% azimuthal regions near an EM module edge suffer some
degradation in identification probability and energy response, but
are typically easily recognizable as electrons.   We will refer to
them as `\edgec ' or `edge' electrons.  The edge region corresponds to about
1.8 cm on either side of the EM module.  
Those electrons identified
in the end calorimeters \cite{ecmw} are labelled `E'.
The end calorimeters have a single full azimuth module and consequently
have no edges.
Dielectron pair samples are denoted CC, \edgecc ,
\edgec \edgec , CE,  \edgece , or EE according to the location of the two
electrons.

\begin{figure}[!hbp]
\epsfig{figure=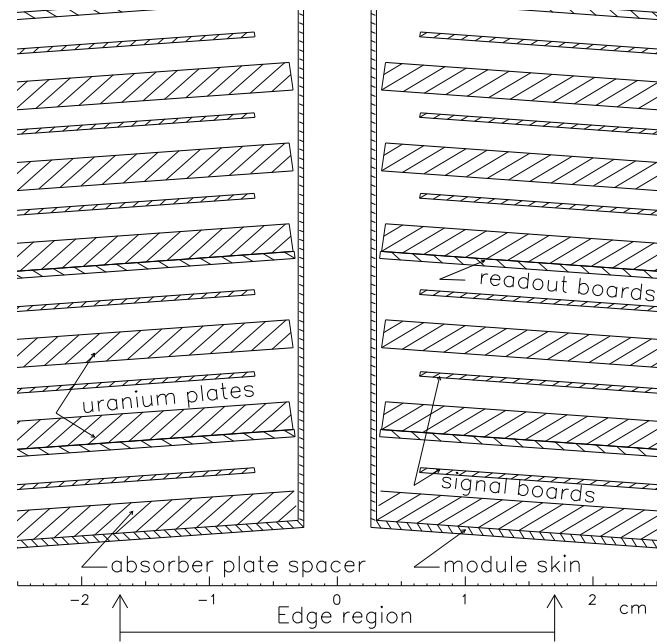,width=8.5cm}
\caption{
Construction of central calorimeter EM modules in the region
near module boundaries.   Signal boards have the electrode pads
for signal collection; readout boards carry traces bringing the signals
to the module ends.
}
\label{fig:ccedge}
\end{figure}

The detailed constitution of the EM calorimeter in the vicinity
of the edges of two modules is shown in
Fig.~\ref{fig:ccedge}.
The mechanical support structure for the modules is provided by
thick stainless steel end plates (not shown); the end
plates of adjacent modules are in contact to form a 32-fold polygonal arch.
The elements of each module are contained within a permeable stainless
steel skin to allow the flow of liquid argon within the cryostat. 
Adjacent module skins are separated by about 6 mm.  
The uranium
absorber plates extend to the skins, so that any electron impinging
upon the module itself will pass through sufficient material to 
make a fully developed EM shower.
Within the gaps between absorber
plates, G10 signal boards are etched on both sides to provide the desired
$\eta-\phi$ segmentation for readout.   The signal boards are coated
on both sides with resistive epoxy and held at a voltage of 2 kV 
to establish the electric field within which ionization drifts
to the signal boards.  The resistive coat is set back from the ends of
the board  by about 3 mm to avoid shorts to the skin.  In the
region of this setback, the electric field fringing causes low
ion drift velocity and thus reduced signal size, but the shower
development is essentially normal as the absorber configuration is
standard.  The hadronic calorimeter modules are rotated in azimuth 
so that the edges of EM and hadronic modules are not aligned.

The directions of electrons and 
their impact point on the calorimeter are determined 
\cite{ccmw1b,ccmw1a} using the central
drift chamber (CDC), located just inside the calorimeter cryostat.  This
chamber has four azimuthal rings of thirty-two modules each.  In each module,
the drift cell is defined with seven axial sense wires and associated field
shaping wires.   The ring 2 and 4 sense wire azimuthal locations
are offset by one half cell from those of rings 1 and 3.   
Half of the sense wires are aligned in azimuth with a calorimeter
edge and the other half are aligned with the center of a calorimeter module.
The drift chamber $z$-coordinate parallel to the beam is measured by
delay lines in close proximity to the inner and outer sense wires of each
module, using the time difference of arrival at the two ends.

\subsection{Triggers}

Triggers 
for the $W$ boson mass analysis, described in more detail in 
Refs. \cite{ecmw,ccmw1b,ccmw1a}, are derived primarily from
calorimetric information.  For the hardware level 1 trigger, 
calorimeter signals
are ganged into $\Delta\eta\times\Delta\phi = 0.2\times0.2$ 
towers in both EM and hadronic sections.   
Energy above a threshold is required for a seed EM tower.
The hardware refines this to include the maximum
transverse energy tower adjacent to the seed, and requires this
combination to exceed a fixed threshold.   The corresponding hadronic
tower transverse energy must not exceed 15\% of the EM tower energy.
The second level trigger refines the information in 
computer processors using a more sophisticated clustering algorithm.
At level 2, the missing transverse energy (\met) components are formed.
The $W$ boson level 2 trigger requires an EM 
cluster and \met ~above a threshold.  The $Z$
boson level 2 trigger requires two EM clusters.  In addition, trigger
requirements are imposed to ensure an inelastic collision, signalled
by scintillators near the beam lines, and require the event to be
collected outside times where beam losses are expected to occur \cite{ecmw}. 
For the offline cuts described below, the triggers are
100\% efficient \cite{ccmw1b,slava}.

\subsection{Data Selection}

The offline data
selection cuts are the same as in the previous \d0 $W$ boson mass
analyses.   The variables used for event selection are:
\begin{itemize}
\item Electron track direction:  The track azimuth 
of a C or \edgec ~electron
is determined from the
CDC track centroid and the reconstructed transverse vertex
position (determined from the drift chamber measurement of tracks). 
We define the axial track center of gravity
in the CDC as $z_{\rm trk}$.
 The track pseudorapidity is then determined 
from the difference between $z_{\rm trk}$
and the EM calorimeter cluster center of gravity.
\item Distance of the electron from calorimeter module edge:
The distance along the front face of the EM
calorimeter module from the module edge is measured by the extrapolation
of the line from the event vertex through the central drift chamber
track centroid.  The azimuthal distance from the module edge is denoted \dedge .
\item Calorimeter energy location:  $\eta_{\rm det}$ is the pseudorapidity
of the EM cluster in the calorimeter, measured from the center of
the detector.  The axial position of
the EM cluster in the EM calorimeter is denoted by $z_{\rm clus}$.
\item Shower shape:  The covariance matrix of energy deposits in
forty lateral and longitudinal calorimeter 
subdivisions and the primary vertex $z$ position are used to define
a chisquare-like parameter, $\xi_{\rm shape}$, 
that measures how closely  a given shower resembles
test beam and Monte Carlo EM showers \cite{hmatrix}.
\item Electron isolation:  the calorimeter energies are used to define
an isolation variable, $f_{\rm iso}=(E_{\rm full}-E_{\rm core})
/E_{\rm core}$, where $E_{\rm core}$ is the energy in the EM
calorimeter within $\mathcal R$=0.2 of the electron direction, 
$E_{\rm full}$ is the energy in the full calorimeter within 
$\mathcal R$=0.4 and $\mathcal{R}=\sqrt{\Delta \eta^2 + \Delta \phi^2}$.
\item Track match significance:  $\sigma^2_{\rm trk}=
(\Delta s / \delta s)^2 + (\Delta\zeta / \delta \zeta)^2 $ measures the
quality of the track match, where $s$ is the $r\phi$ coordinate
and $\zeta$ is the $z$ coordinate for the central
calorimeter or radial coordinate for the end calorimeter.   
$\Delta s$ and $\Delta \zeta$ are the differences between track projection
and shower maximum coordinates in the EM calorimeter, and
$\delta s$ and $\delta\zeta$ are the corresponding errors 
\cite{ecmw,ccmw1b}.
\item EM fraction:  the fraction, EMF, of energy within a cluster 
that is deposited in the EM portion of the calorimeter.
\item Electron likelihood:  a likelihood variable, $\lambda_4$, based upon a
combination of EMF, $\sigma_{\rm trk}$, $dE/dx$ in the CDC,
and $\xi_{\rm shape}$ \cite{fourlikelihood}.
\item Kinematic quantities:  the transverse momenta of electrons,
neutrinos, and the $W$ or $Z$ bosons are  denoted
$p_T(e), p_T(\nu), p_T(W)$ or $p_T(Z)$.  The \ptnu ~is determined
from the missing transverse energy in the event, as discussed below. 
The invariant mass of two electrons is denoted by $m_{ee}$.  
\end{itemize}
The requirements for central and end electrons are given
in Table~\ref{tab:offlineselect}.

\begin{table}
\caption{\label{tab:offlineselect}
Offline selection criteria for central
and end electron candidates.
}
\begin{ruledtabular}
\begin{tabular}{c|c|c} 
Variable &  Central Electron & End Electron \\ \hline
$|\eta_{\rm det}|$ & $\leq 1.1$& $1.5 - 2.5$\\ 
$\xi_{\rm shape}$  & $\leq 100$& $\leq 200$ \\ 
$\sigma_{\rm trk}$ & $\leq 5$  & $\leq 10$   \\ 
EMF                & $\geq 0.90$ & $\geq 0.90$\\ 
$f_{\rm iso}$      & $\leq 0.15$ & $\leq 0.15$ \\ 
$\lambda_4$        & --        & $\leq 4.0$   \\ 
$|z_{\rm clus}|$   & $\leq 108$ cm & --      \\ 
$|z_{\rm trk}|$    & $\leq 80$ cm & --       \\ 
\end{tabular}
\end{ruledtabular}
\end{table}

The selection criteria for the $W$ and $Z$ boson event samples are
given in Table~\ref{tab:selectioncuts}.  
Non-edge electrons are defined as those with \dedge /$d_{\rm mod} \geq 0.1$,
where $d_{\rm mod}$ is the full width of the module in azimuth. Edge electrons
are required to have \dedge /$d_{\rm mod}$ $< 0.1$.
For the $Z$ boson sample with two electrons in the central
calorimeter, both are required to
have good tracks in the drift chamber 
(\textit{i.e.} passing the $\sigma_{\rm trk}$ requirement)
if either of them is in a
central calorimeter edge region; if both are non-edge, only one electron is
required to have a good track. 
For $Z$ boson samples with one electron
in the end calorimeter, the end electron must have
a good track, while the central electron is required to have a good
track only if the electron is in the edge region.

\begin{table}
\caption{\label{tab:selectioncuts}
 Event selection criteria for $W$
and $Z$ boson samples.}
\begin{ruledtabular}
\begin{tabular}{c|c|c}
Variable &  $W$ boson sample~~ & $Z$ boson sample~~ \\ \hline
$p_T(e{\rm ~central})$ & $\geq 25$ GeV & $\geq 25$ GeV \\
$p_T(e{\rm ~end})$   &  --             & $\geq 30$ GeV \\
$p_T(\nu)$ & $\geq 25$ GeV & --                  \\
$p_T(W)$   & $\leq 15$ GeV & --                  \\
$m_{ee}$   & --            & 60 -- 120 GeV    \\
$|z_{\rm vtx}|$ & $\leq 100$ cm  & $\leq 100$ cm \\
\end{tabular}
\end{ruledtabular}
\end{table}

With these selections, we define three $W$ boson samples and 
six $Z$ boson samples, differentiated by whether the electrons used
are C, \edgec , or E.   The numbers of events selected in each sample
are given in Table~\ref{tab:samplesize}.

\begin{table}
\caption{\label{tab:samplesize}
Event sample sizes.}
\begin{ruledtabular}
\begin{tabular}{cr|cr} 
$W$ boson sample &  No. events~~  & $Z$ boson sample & No. events~~ \\ \hline
C & 27,675~~ & CC & 2,012~~ \\
$\widetilde{\rm C}$ & 3,853~~ & $\widetilde{\rm C}$C & 470~~ \\
E & 11,089~~ & $\widetilde{\rm C}$$\widetilde{\rm C}$ & 47~~ \\
~ & ~ & CE & 1,265~~ \\
~ & ~ & $\widetilde{\rm C}$E & 154~~ \\
~ & ~ & EE & 422~~ \\
\end{tabular}
\end{ruledtabular}
\end{table}

\subsection{Experimental method}

The experimental method used in this work closely resembles that
of previous \d0 $W$ boson mass measurements.
We compare distributions from the $W$ and $Z$ boson samples with
a set of templates of differing mass values, 
prepared using a fast Monte Carlo program that 
simulates vector boson production and decay, and incorporates the
smearing of experimentally observed quantities using distributions
derived from data. 
The variables used for the $W$ boson templates are the transverse mass,
\begin{equation}
m_T = \sqrt{2p_T(e)p_T(\nu)[1-\cos(\phi_e-\phi_\nu)]} ,
\end{equation}
and the transverse momenta of the electron and neutrino, $p_T(e)$ 
and $p_T(\nu)$.   The three distributions depend on a common
set of detector parameters, but with different functional
relationships,
so that the measurements from the three distributions are
not fully correlated.   
As discussed in Ref. \cite{ecmw}, the \mt ~distribution is affected most
by the hadronic calorimeter response parameters, whereas the 
\pte ~distribution is mainly broadened by the intrinsic 
\ptw ~distribution, and the 
\ptnu ~distribution is smeared by a combination of both effects.
The $Z$ boson template variable is the
invariant mass, $m_{ee}$.

The observed quantities used for $W$ boson reconstruction are 
$p_T(e)$ and the recoil transverse momentum,
$\vec u_T = \Sigma_i E_{Ti} \hat n_i$, where $\hat n_i$ is the unit vector
pointing to calorimeter cell $i$, and the sum is over all calorimeter
cells not included in the electron region.  The electron energy in
the central calorimeter is summed over a $\Delta\eta\times \Delta\phi$
region of $0.5\times0.5$ centered on the most energetic calorimeter
cell in the cluster.  
Note that this region spans 2.5 modules in azimuth, so always contains several
module edges irrespective of the electron impact point.
For the end calorimeter, the electron energy sum
is performed within a cone of radius 20 cm (at shower maximum), 
centered on the
electron direction.  In both cases energy from the EM calorimeter
and first section of the hadron calorimeter is summed.

The neutrino transverse momentum in $W$ boson decays is taken to be 
$\vec p_T(\nu) = -\vec p_T(e) - \vec u_T$.  The components of
$\vec u_T$ in
the transverse plane are most conveniently taken as
$u_\parallel = \vec u_T \cdot \hat e$ and 
$u_\perp = \vec u_T \cdot (\hat e\times\hat z)$, where $\hat e$ 
($\hat z$) is the electron (proton beam) direction.

The momentum $\vec p(ee) = \vec p(e_1)+\vec p(e_2)$
and the dielectron invariant mass define the  dielectron 
system for the $Z$ boson sample.  The dielectron transverse
momentum is expressed  in components 
along the inner bisector axis $\hat \eta$ of the two
electrons, and the transverse axis $\hat \xi$ perpendicular to 
$\hat \eta$.

The data are compared with each of the templates
in turn and a likelihood parameter $\mathcal L$ is calculated.
The set of likelihood values at differing boson masses
and fixed width is fitted to find the 
maximum value, 
corresponding to the best measurement of the mass.  Statistical errors
are determined from the masses at which $\ln \mathcal{L}$ decreases by one-half
unit from this maximum.

\subsection{Monte Carlo production and decay model}

The production and decay model is taken to be the same as for
the earlier measurements \cite{ecmw,ccmw1b,ccmw1a}.  
The Monte Carlo production cross section is based upon a 
perturbative calculation \cite{ladinskyyuan} which depends on
the mass, pseudorapidity, and transverse momentum of the produced boson,
and is convoluted with the MRST parton distribution
functions \cite{mrst}.
We use the mass-dependent Breit-Wigner function \cite{ccmw1b} with measured
total width parameters $\Gamma_W$ and $\Gamma_Z$ to 
represent the line shape of the vector bosons.  The line shape
is modified by the relative parton luminosity as a function
of boson mass, due to the effects of the parton distribution
function.  The parameter $\beta$ in the parton luminosity function 
$ \mathcal{L}_{q\overline q} = e^{-\beta m_{ee}}/m_{ee}$ 
is taken from our previous studies \cite{ecmw,ccmw1b}.

Vector boson decays are simulated using matrix
elements which incorporate the appropriate helicity states of the
quarks in the colliding protons and antiprotons.   Radiative decays of
the $W$ boson are included in the Monte Carlo model \cite{ccmw1b} based on the
calculation of Ref. \cite{behrends}.  Decays of the $W$ boson into
$\tau \nu$ with subsequent $\tau \rightarrow e\nu\overline\nu$ 
decays are included
in the Monte Carlo, properly accounting for the $\tau$ polarization 
\cite{ccmw1b}.

\subsection{Monte Carlo detector model}

The Monte Carlo detector model employs a set of parameters for
responses and resolutions taken from the data \cite{ccmw1b}.  
Here we summarize these
parameters and indicate which are re-evaluated for the edge electron
analysis. 

The observed electron energy response is taken to be of the form
\begin{equation}
\label{responseenergy}
E^{\rm meas}=\alpha E^{\rm true} +\delta .
\end{equation}
The scale factor $\alpha$ that corrects the response relative to test
beam measurements is determined using fits to the $Z$ boson
sample; for the C electrons, $\alpha=0.9540\pm 0.0008$.
The energy offset parameter $\delta$ correcting for effects of
uninstrumented material before the calorimeter is found from fits to
the energy asymmetry of the two electrons from $Z$ bosons,
and from fits to $J/\psi\rightarrow e^+e^-$ and 
$\pi^0\rightarrow \gamma \gamma \rightarrow (e^+e^-)(e^+e^-)$ decays.  
For C electrons, $\delta=-0.16^{+0.03}_{-0.21}$.  There
is an additional
energy correction (not shown in Eq.~\ref{responseenergy}) 
that contains the effects of the
luminosity-dependent energy depositions within the electron window
from underlying events, and also corrects for the effects of noise and
zero suppression in the readout.  This correction is 
made using observed energy depositions in $\eta-\phi$ control regions 
away from electron candidates.
We discuss the modification of the energy response parametrization
for \edgec ~electrons below.

The electron energy resolution is taken as 
\begin{equation}
\label{resolutionenergy}
{\sigma_E\over E}={s\over\sqrt E} \oplus c \oplus {n\over E} ,
\end{equation}
where $\oplus$ indicates addition in quadrature.  The sampling term
constant $s$ is fixed at the value obtained 
from test beam measurements, and the noise term
$n$ is fixed at the value obtained from the observed uranium and 
electronics noise distributions in the
calorimeter.  The constant term $c$ is fitted from the observed
$Z$ boson line shape.  
The parameter values for C electrons \cite{ccmw1b} are
$s=0.135$ (GeV$^{1/2}$), 
$c=0.0115^{+0.0027}_{-0.0036}$, and $n=0.43$ GeV.
The resolution parameters are re-evaluated
for \edgec ~electrons below.

The transverse energy is obtained from the observed energy using
$E_T = E\sin\theta$, where the polar angle is obtained as indicated
in Sec. IIC, with the errors taken from the measurements of electron tracks
in $Z$ boson decays.

The efficiency for electron identification depends on the amount
of recoil energy, $u_\parallel$, along the electron direction.
We take this efficiency to be constant for $u_\parallel < u_0$
and linearly decreasing with slope $s_0$ for $u_\parallel > u_0$.
The parameters of this model for the efficiency are 
determined by superimposing Monte Carlo electrons
onto events from the $W$ boson signal sample with the electron removed, 
and then subjecting the event to
our standard selection cuts.  
For non-edge electrons, $u_0=3.85$ GeV and $s_0=-0.013$ GeV$^{-1}$; these
parameters are strongly correlated \cite{ccmw1b}.
Since the properties of electrons
in the edge region are different from those in the non-edge region, we
re-examine this efficiency below for the \edgec ~sample.

The unsmeared recoil transverse energy is taken to be
\begin{equation}
\vec u_T=-(R_{\rm rec}~\vec q_T) 
-\Delta u_\parallel ~\hat p_T(e) +
\alpha_{\rm mb}~\hat m  ,
\end{equation}
where $\vec q_T$ is the generated $W$ boson transverse momentum; 
$R_{\rm rec}$ is the response of the calorimeter to recoil
(mostly hadronic) energy; $\Delta u_\parallel$ is a luminosity-
and $u_\parallel$-dependent correction for energy flow into the
electron reconstruction window; $\alpha_{\rm mb}$ is a correction
factor that adjusts the resolution to fit the data, and is roughly the
number of additional minimum bias events overlaid on a $W$ boson event;
and $\hat m$ is the unit vector in the direction of the 
randomly distributed minimum bias
event transverse energy.
The response parameter is parametrized as 
$R_{\rm rec} = \alpha_{\rm rec}+\beta_{\rm rec}\log q_T$ and is
measured using the momentum balance 
in the $\hat \eta$ (dielectron bisector) direction  for the $Z$ boson
and the recoil system.  The $\Delta u_\parallel$ parameter due to
recoil energy in the electron window is similar
to the corresponding correction to the electron energy, but is
modified to account for readout zero-suppression effects.
The recoil response is due to energy deposited over all the
calorimeter, and thus is not expected to be modified for the 
\edgec ~electron analysis.

The recoil transverse energy resolution is parameterized as a Gaussian response
with $\sigma_{\rm rec}= s_{\rm rec}\sqrt{u_T}$, modified by the
inclusion of a correction for luminosity-dependent 
event pileup controlled by the
$\alpha_{\rm mb}$ parameter introduced above.  These parameters are fit
from the $Z$ boson events using the spread of the $\hat \eta$ component
of the momentum balance of the dielectron-recoil system.
Since the $s_{\rm rec}$ term grows with $p_\eta(ee)$ while the
$\alpha_{\rm mb}$ term is independent of $p_\eta(ee)$, the two
terms can be fit simultaneously.   The recoil resolution parameters
are not expected to differ for the C and \edgec ~samples.

\section{Background determination}
\label{background}

As noted above,
the $W\rightarrow\tau\nu\rightarrow e\nu\overline\nu \nu$ background
is included in the Monte Carlo simulation.  Because of the branching ratio
suppression and the low  electron momentum, 
this background is small (1.6\%
of the $W$ boson sample).  
The remaining estimated backgrounds 
discussed in this section are added to the Monte Carlo event
templates for comparison with data. 

The second background to the \edgec ~$W$ boson
sample arises from $Z\rightarrow e^+e^-$ events in which one electron is
misreconstructed or lost.  It is taken to be the same as for the C sample,
($0.42 \pm 0.08$)\%,
since the missing electron is as likely to be an edge electron
for both C and \edgec ~samples.
Small differences in the shape of this background in the case where
one $Z$ boson electron falls in the edge region give negligible modification to
the final $W$ boson mass determination.

\begin{figure}[!hbp]
\epsfig{figure=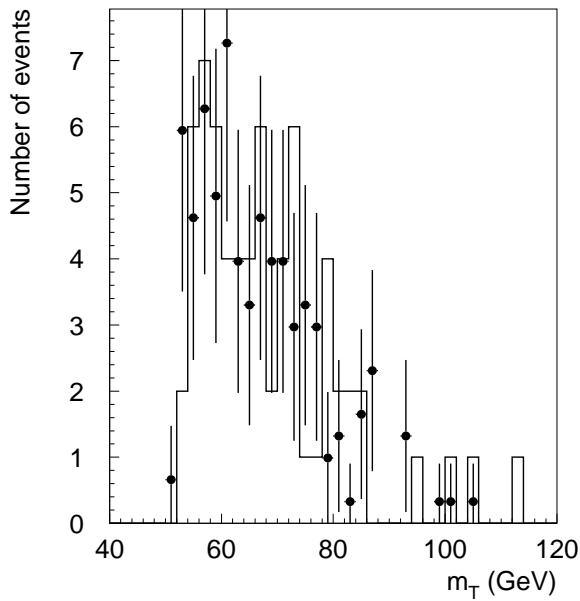,width=8.5cm}
\caption{
Comparison of transverse mass distributions for background
events to $W$ bosons for C (points with error bars) 
and \edgec ~(solid histogram).
The two distributions are normalized to the same number of events.
}
\label{fig:wqcdback}
\end{figure}

The third background for the $W$ sample is due to QCD multijet events in
which a jet is misreconstructed as an electron.   This background is
estimated by
selecting events with low \met ~using a special trigger which
is dominated by QCD jet production.  
For events with \met ~$<$ 15 GeV, we compare the number of events with 
`good' and `bad' electrons.  Good electrons are required to pass all
standard electron identification cuts, whereas bad electrons have 
track match selection cut $\sigma_{\rm trk} > 5$
and require $\xi_{\rm shape} > 100$.  We
assume that the probability for a jet to be misidentified as 
an electron does not
depend on  \met , and determine it for both C and 
\edgec ~samples.   The \mt ~distributions for both
C and \edgec ~samples are shown in 
Fig.~\ref{fig:wqcdback}.  
Here, and for the \pte ~and \ptnu ~distributions, the C and \edgec ~samples
are statistically indistinguishable;
the fraction of background events in the non-edge $W$ boson sample is 
($1.3\pm 0.2$)\%, whereas for the edge sample it is
($1.5\pm 0.2$)\%.  We use the QCD multijet background distribution from
the C sample \cite{ccmw1b} for the \edgec ~analysis.

The background for the $Z$ boson sample is 
composed of QCD multijet events with
jets misidentified as electrons.   We evaluate this background from
the dielectron mass distributions with two 
`bad' electrons, one in the edge region and one in the 
non-edge region.   We find an exponentially decreasing
shape of the background as a function of $m_{ee}$ with a slope
parameter of $-0.064 \pm 0.022$ GeV$^{-1}$ 
for the \edgecc ~sample, to be compared with
a slope of $-0.038 \pm 0.002$ GeV$^{-1}$ for the CC sample, so we use
different background shapes for the two samples.  The fraction of
events in the mass region $70 \leq m_{ee} \leq 110$ GeV is
($3.7\pm3.6$)\% for the \edgecc ~sample and
($2.2\pm1.3$)\%  for CC.   
The \edgece ~$Z$ boson background 
is statistically indistinguishable from the CE $Z$ boson sample,
so we use the background distribution determined in Ref. \cite{ecmw}
for the \edgece ~$Z$ boson analysis.

\section{Edge electron energy response and resolution}
\label{response}

\subsection{Determination of edge electron response and resolution
parameters}

The thirty-two central calorimeter modules are about 18 cm wide in the
$r\phi$ direction at the shower maximum.  
Thus the edge regions defined above are about 1.8 cm wide.
The Moli\`ere radius
$\lambda_M$  in the composite material of the
\d0 calorimeter is 1.9 cm.   Since electrons deposit 90\% of their
energy in a circle of radius 1 $\lambda_M$ (and about 70\% within 0.5
$\lambda_M$), the choice was made in all previous \d0 analyses using
central electrons to make a fiducial cut excluding electrons within
the 10\% of the module nearest the edge.  
As noted in Section \ref{experiment}, we expect that
showers will develop normally over the portion of the central
calorimeter module edges where energy can be recorded, but that the actual
energy seen may be degraded.  In this section we motivate modified
edge electron energy response and resolution functions, and describe the
determination of the associated parameters. 

\begin{figure}[!hbp]
\epsfig{figure=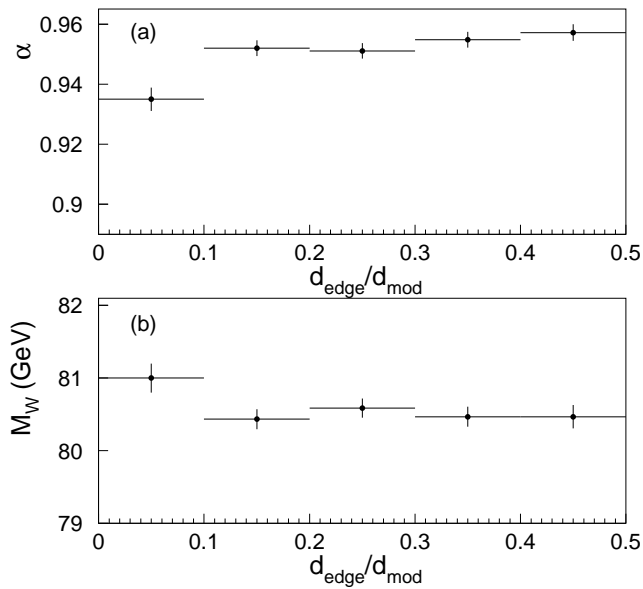,width=8.5cm}
\caption{
Distributions for \edgec ~samples as a function of the ratio of the
electron impact distance \dedge ~from the module edge to the total
module width, $d_{\rm mod}$:  (a) the
fitted scale factor $\alpha$, and (b) the fitted $W$ boson mass 
using the appropriate scale factor for each \dedge ~bin. The errors
are statistical only.}
\label{fig:scaleandmwsimple}
\end{figure}

A naive modification for the electron energy response and resolution
parametrization would use the same forms 
(Eqs. \ref{responseenergy} and \ref{resolutionenergy}) 
employed for the non-edge analyses
with changed values for some of the parameters.   Since the primary
effect expected as the
distance, \dedge , of an electron from the module edge varies
is the loss of some signal, we might 
consider modified values for the parameter $\alpha$.  
Figure~\ref{fig:scaleandmwsimple}(a)
shows the result of a fit for the scale factor 
$\alpha$ in a sample of $Z$ boson events in which one electron is in
a non-edge region, as a function of the position of the second electron.
A clear reduction in
$\alpha$ is observed in the edge bin.  When the value appropriate for
each bin in \dedge ~is used in the analysis for
the $W$ boson mass, we see a significant deviation of $M_W$ in the edge
bin, as shown in Fig.~\ref{fig:scaleandmwsimple}(b).  
Modifying both $\alpha$ and 
the parameter
$c$ in the resolution function does not improve the agreement for
$M_W$ in different regions.
We conclude that this simple modification of energy response is inadequate.   

\begin{figure}[!hbp]
\epsfig{figure=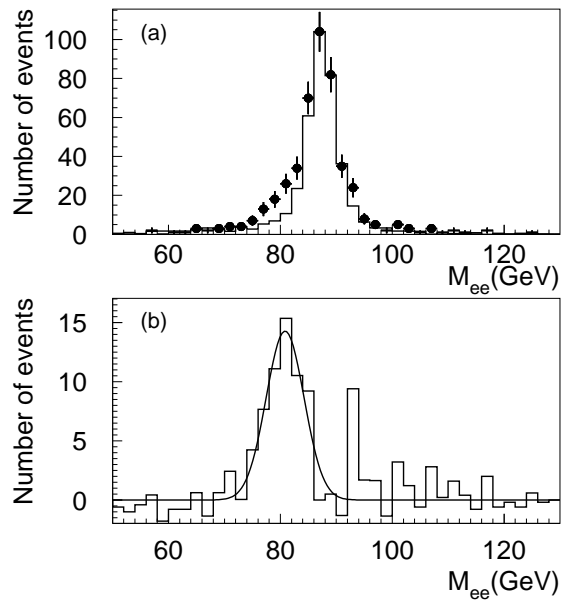,width=8.5cm}
\caption{
(a) Dielectron mass distributions for CC and 
\edgecc ~samples, with the CC distribution scaled to give
the same peak value as for the \edgecc ~distribution.  The solid histogram
is for the CC $Z$ bosons and the points are for the \edgecc ~$Z$ bosons.
(b) The difference
between \edgecc ~and normalized CC samples.  The curve is a Gaussian fit; 
no backgrounds are included in the fit to the difference.
}
\label{fig:zmasscomparison}
\end{figure}

Insight into the appropriate modification to the electron response and
resolution can be gained by comparing the $Z$ boson mass distributions
for the case of both electrons in the non-edge region (CC)
to that when one electron is in the edge region and the other is non-edge
(\edgecc ).  
Figure~\ref{fig:zmasscomparison}(a)
shows both distributions (before any energy response scaling), 
normalized to the same peak amplitude.
The \edgecc ~distribution agrees well with the CC sample at mass values
at and above the peak in the mass distribution, but exhibits an excess on the
low mass side.   When the CC distribution is subtracted from the
\edgecc ~distribution, the result is the broad Gaussian shown in 
Fig. 5(b), centered at about 95\% of the mass value for the CC sample.

The data suggest a parametrization of edge electron 
response in which
there are two components.  The first is a Gaussian function with the same
response and resolution parametrizations
as for the non-edge electrons, for a fraction (1-\fedge ) of
the events:
\begin{equation}
\label{responsenonedge}
E^{\rm meas}=\alpha E^{\rm true} + \delta
\end{equation}
\begin{equation}
\label{resolutionnonedge}
{\sigma_E\over E}={s\over\sqrt E} \oplus c \oplus {n\over E} ,
\end{equation}
and the second is a Gaussian with
reduced mean and larger width to describe the lower energy subset
of events.   Guided by the data, we take 
the same functional description for the response and resolution parameters
for a fraction \fedge ~of events:
\begin{equation}
\label{responseedge}
E^{\rm meas}=\widetilde{\alpha} E^{\rm true} + \widetilde{\delta}
\end{equation}
\begin{equation}
\label{resolutionedge}
{\sigma_E\over E} = {\widetilde{s}\over\sqrt E} \oplus \widetilde{c} \oplus 
{\widetilde{n}\over E} .
\end{equation}
The parameters in Eqs.~\ref{responsenonedge} and \ref{resolutionnonedge}
denoted without a tilde are those from the previous
non-edge $W$ boson mass analysis \cite{ccmw1b}.  Those with the tilde
in Eqs.~\ref{responseedge} and \ref{resolutionedge}
are in principle new parameters for the fraction \fedge ~of edge electrons with
reduced signal response.

The modified response is characterized by a reduction in
the average energy seen for a fraction of the edge electrons and on average 
a reduced EMF for edge electrons.  
A potential explanation for the energy reduction as being due to electrons
that pass through the true crack between EM calorimeter modules is not 
satisfactory.
In this case the energy missing in the EM section would be recovered in
the hadronic calorimeter modules giving the correct full electron energy.
(We note that there is only a 14\% increase in the number of $W$ boson 
electrons (\textit{c.f.} Table~\ref{tab:samplesize})
when the azimuthal coverage is increased by 25\% by including
the edge region, indicating
that some electrons in the true intermodular crack are lost from
the sample.)

A more plausible hypothesis is that
the electrons in the edge region shower in the
EM calorimeter normally, but for the subset of electrons which pass
near the module edge, 
the signal is reduced due to the smaller electric drift field in
the edge region.  In this case too, the average EMF is reduced 
due to the loss of some EM signal, but the overall energy is lowered as well.
This picture of the energy response agrees with the observed behavior
seen in Fig. 5.
Our model is probably oversimplified, since even within 
the edge region there can be a range
of distances between shower centroid and the module edge where the
electric field is most affected, leading to variable amounts of lost signal.   
The distribution of Fig. 5(b) however indicates that a
single extra Gaussian term in the response suffices to explain the 
data at the present level of statistical accuracy.  
We speculate that the convolution over impact position
contributes to the rather large width of the
lower energy Gaussian term, relative to that for the full energy
Gaussian.

The representation above for edge electron response and resolution
introduces six potential new parameters: \alphaedge ,
\deltaedge , \sedge , \cedge , \nedge ~and \fedge .
We expect \nedge ~= $n$ since the electronics noise
should be unaffected near the edge of a module.   

Since there is no difference in the amount of material before
the calorimeter, we would expect that \deltaedge ~$= \delta$.
The determination of $\delta$ can be made from the $Z$ boson sample data.
For the form of the energy response function adopted above, 
the observed $Z$ boson invariant mass, $m_{ee}$, should be
\begin{equation}
m_{ee} = \alpha M_Z +  \mathcal{F}_Z \delta 
\end{equation}
in the case that $\delta \ll E(e_1)+E(e_2)$.  Here, $M_Z$ is the true
$Z$ boson mass taken from LEP measurements \cite{ewwkgp} 
($M_Z=91.1875$ GeV),  
$ \mathcal{F}_Z=[E(e_1)+E(e_2)] (1-\cos\omega)/m_{ee}$, 
and $\omega$ is the opening angle between the two electrons $e_1$
and $e_2$.   
Fitting the dependence of $m_{ee}$ on $ \mathcal{F}_Z$  
\cite{ccmw1b} gives $\delta$.   We find that the $ \mathcal{F}_Z$ dependence 
for the \edgecc ~$Z$ boson sample is consistent ($\chi^2 = 8.9$ for 9 degrees
of freedom) with that for the CC $Z$ boson sample, and thus take
\deltaedge ~$= \delta$.

\begin{figure}[!hbp]
\epsfig{figure=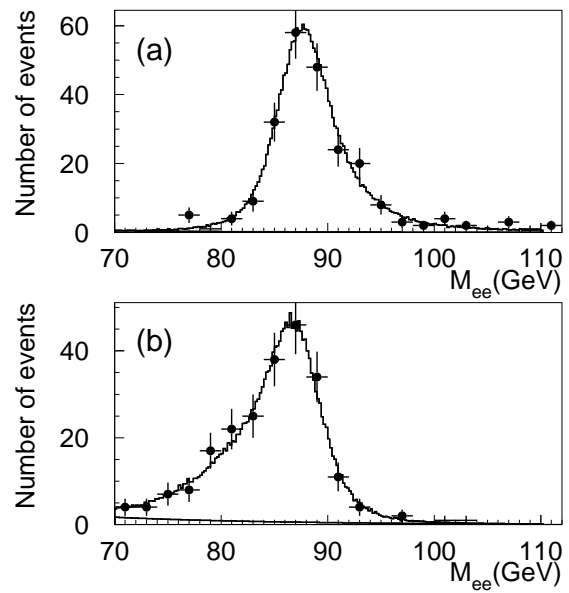,width=8.5cm}
\caption{
$Z$ boson mass distributions for (a)
edge electrons with $E_T > 41$ GeV and (b)
edge electrons with $E_T < 41$ GeV.  The histograms are the 
best fit distributions from the Monte Carlo.  The curve at
the bottom of (b) represents the background.
}
\label{fig:samplingterm}
\end{figure}

We argued above that, because the structure of the absorber plates
extends well past the region where the high voltage plane ends, we
would expect the same sampling constants in edge and non-edge regions.
We check this hypothesis by dividing the \edgecc ~$Z$ boson sample
into two equally populated bins of edge electron energy,
$E_{e} < 41$ GeV and $E_{e} > 41$ GeV, for
which the mean energies are 36 and 47 GeV respectively.  
Using the non-edge value
of $s$ for both subsamples, we show in 
Fig.~\ref{fig:samplingterm}  
the $Z$ boson mass distributions and the Monte Carlo 
expectation for the best template fit described in more
detail below.
We find the fitted $Z$ boson masses are
$91.10 \pm 0.32$ GeV ($E_{e} < 41$ GeV) with $\chi^2 = 4.5$ for 14
degrees of freedom and
$91.06 \pm 0.27$ GeV ($E_{e} > 41$ GeV)
with $\chi^2 = 12$ for 16 degrees of freedom.
The consistency and goodness of fit leads us to take \sedge ~$= s$.

\begin{figure}[!hbp]
\epsfig{figure=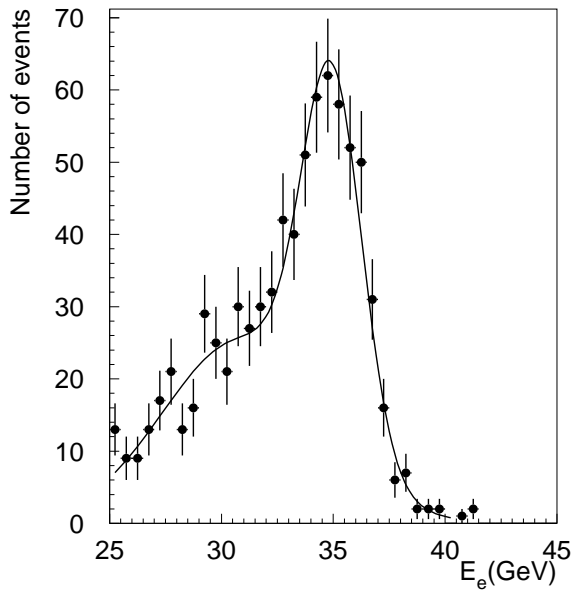,width=8.5cm}
\caption{
Monte Carlo simulation of the energy response
function for 40 GeV electrons in the edge region.  The points represent
the Monte Carlo data and a fit using the parametrization of
Eqs.~\ref{responsenonedge}--\ref{resolutionedge} is given by the curve.
}
\label{fig:mcresponse}
\end{figure}

We simulate the response of the calorimeter to electrons
in the edge region, using the \textsc{geant} \cite{geant} program with
all uranium plates and argon gaps included.  The simulation lacks
some details of the actual calorimeter, including some of the
material between calorimeter modules, and contains an incomplete simulation
of the detailed resistive coat pattern on the signal readout
boards.   The resulting distribution of energy for 40 GeV electrons
impacting upon the edge region of the calorimeter modules 
is shown in Fig.~\ref{fig:mcresponse}.  The Monte Carlo
distribution closely resembles that seen in the data, with a fraction
of events showing a broad Gaussian with 
lower average response than the main component
of electrons.   Within the imperfect simulation of calorimeter details, 
the agreement with the data is good.   The Monte Carlo distribution
can be well fit with the same functional form  
(Eqs.~\ref{responsenonedge}--\ref{resolutionedge}) used for the data.

\begin{figure}[!hbp]
\epsfig{figure=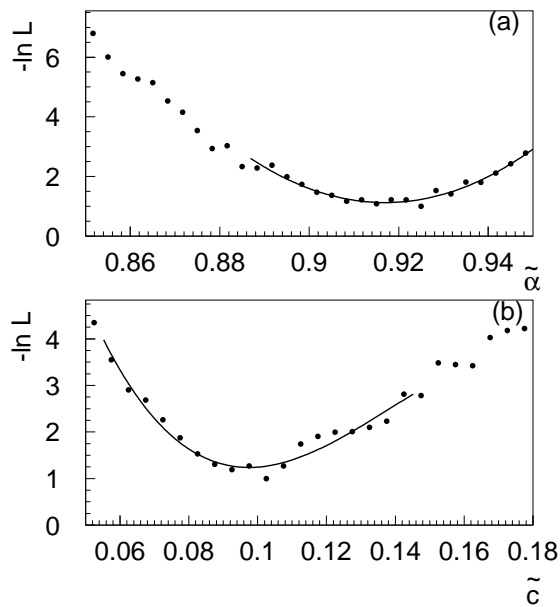,width=8.5cm}
\caption{
(a) Fits to $\widetilde\alpha$ with edge electron parameters $\widetilde c$ and
$\widetilde f$ fixed near their optimum values; 
(b) fits to $\widetilde c$ with edge electron parameters $\widetilde \alpha$ and
$\widetilde f$ fixed near their optimum values.  The curves are
best-fit parabolas.
}
\label{fig:alphacfit}
\end{figure}

\begin{figure}[!hbp]
\epsfig{figure=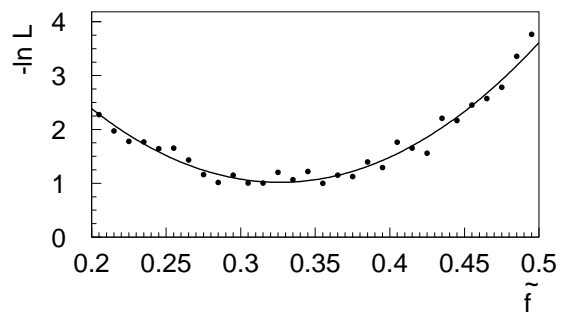,width=8.5cm}
\caption{
Fits to $\widetilde f$ with edge electron parameters $\widetilde \alpha$ and
$\widetilde c$ fixed at their optimum values. The curve
is a best fit parabola.
}
\label{fig:fedgefit}
\end{figure}

Thus, we conclude that for the \edgec ~electrons, we must introduce
only three new parameters \alphaedge , \cedge
~and \fedge .  In principle, we expect that these parameters may be
correlated.  Our fitting procedure is to first fit the \edgecc
~$Z$ boson mass distribution with uncorrelated free parameters 
\alphaedge , \cedge ~and \fedge .    We use the resultant value
\fedge ~= 0.31 as input to a two-dimensional binned likelihood fit
of the templates to the data created by the Monte Carlo, varying
both \alphaedge ~and \cedge .  
The two-dimensional contours show
that the correlation between \alphaedge ~and \cedge ~is very small.
Thus in the vicinity of the maximum likelihood in the two-dimensional 
fit, we can fit one-dimensional 
distributions for each parameter separately.  The one-dimensional 
fits for \alphaedge ~and \cedge ~are repeated iteratively
after modifying the other parameter;  the
process converges after one iteration.  The results
of these fits, shown in 
Fig.~\ref{fig:alphacfit}, give
\alphaedge ~$=0.912\pm 0.018$ and \cedge ~$=0.101^{+0.028}_{-0.018}$.
For these best fit \alphaedge ~and \cedge , we make a one-dimensional
fit for \fedge ~as shown in
Fig.~\ref{fig:fedgefit}
and find \fedge ~$=0.346 \pm 0.076$.

To verify that the non-edge scale factor $\alpha$ and
the narrow Gaussian width from the non-edge electrons 
are indeed appropriate for the fraction (1-\fedge ) of edge electrons
represented with standard response, we perform a fit
to the \edgecc ~$Z$ boson sample in which both narrow and
wide Gaussian parameters are allowed to vary.  The resulting
values for $\alpha$ and $\sigma_E$ for the narrow Gaussian are
consistent with those obtained in the non-edge analysis \cite{ccmw1b}.

We also look for a dependence of the response parameters
on the electron selection variables EMF, $f_{\rm iso}$, $\xi_{\rm shape}$
and $\sigma_{\rm trk}$ by breaking the $Z$ boson sample into bins
of each of these variables and fitting for the edge fraction \fedge ~
within each bin.  No significant variations are seen.  The largest
is a one-standard-deviation slope in the fitted \fedge ~vs EMF distribution,
and we examine the effect of this small dependence as a cross-check below.

\begin{figure}[!hbp]
\epsfig{figure=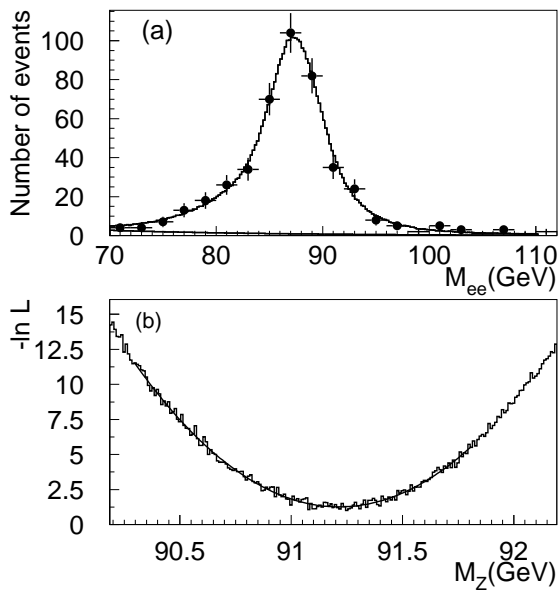,width=8.5cm}
\caption{
(a) Best fit to the 
\edgecc ~$Z$ boson mass distribution using
the parametrization discussed in the text for edge electron response
and resolution.  The lower curve is the expected background.  
(b) The likelihood function as a function of
hypothesized $Z$ boson mass.
}
\label{fig:zmassfit}
\end{figure}

The resulting likelihood
fit to the \edgecc ~$Z$ boson mass using the parametrization
given above is shown in 
Fig.~\ref{fig:zmassfit}.  For this fit, a set of $Z$ boson 
events is weighted in turn to correspond to templates
of $Z$ boson samples spaced at 10 MeV intervals.
The best fit yields $M_Z=91.20\pm0.20$ GeV, with a 
$\chi^2 = 10.4$ for   19 degrees of freedom.
The fitted $Z$ boson mass agrees very well with the input $Z$ boson mass
from LEP \cite{ewwkgp} used in establishing the
parameters \alphaedge , \cedge ~and \fedge.  The small, statistically
insignificant, deviation from the input value occurs since we use the
values of parameters $\alpha$ and $c$ from Ref. \cite{ccmw1b} and
not those which give the absolute minimum $\chi^2$ when these
parameters are varied in the \edgecc ~analysis.

We also investigate alternate parametrizations for the edge electrons 
involving a Gaussian-like function with energy-dependent width
or amplitude.   If we adopt the requirement that such parametrizations
add no more than three new parameters, as for our choice above, we
find such alternatives to be inferior in their ability to represent
the $Z$ boson mass distribution.

\subsection{Cross checks for edge electron response and resolution
parameters}

We noted above that the fraction \fedge ~of 
reduced response electrons in the edge
region displays some dependence upon the
fraction of the total energy seen in the EM section.   Thus our
fitted parameters have been averaged over a range of EMF values.
To check that this averaging is acceptable, we perform analyses
separately on approximately equal-sized subsets of 
events with low and high EMF fractions (EMF $<0.99$ and EMF $>0.99$), for
both the \edgecc ~$Z$ and \edgec ~$W$ boson samples. 
(Values of EMF $>$ 1 are possible due to negative noise fluctuations
in the hadron calorimeter energy.)
For the \edgecc
~$Z$ boson sample, no EMF requirement is made on the C electron.   Since 
the values of the $Z$ boson mass in the
low and high EMF CC $Z$ boson sample
subsets differ slightly, and the 
energy scale parameter $\alpha$ for non-edge electrons is used in
the edge electron response function, we determine the appropriate
$\alpha$'s  for the two EMF ranges of the CC data separately.
The relative change for the scale factor $\alpha$ for the low EMF
non-edge electrons is $-0.17$\%, and for the high EMF selection is $+0.32$\%.
Using these modified values for $\alpha$, we fit the edge electron
parameters \alphaedge , \cedge ~and \fedge ~for each subrange
separately.   Using these results, we create templates using the
modified parameters and fit for the $W$ and $Z$ boson masses 
in both subranges.   The transverse mass distribution was
used to obtain $M_W$.
Table~\ref{tab:lohiemf} shows the fitted parameters and
the resultant mass fits for low and high EMF  subsets.
The $W$ and $Z$ masses agree between the two subsets; the difference
in the fitted $Z$ boson mass between the high and low EMF subsets is
$-0.47 \pm 0.39$ GeV, and for the $W$ boson mass is
$0.62 \pm 0.45$ GeV.  As expected, the fraction \fedge ~is 
larger for the low EMF 
subset, and the width parameter of the Gaussian resolution \cedge ~is larger.
The errors quoted are statistical only; we estimate that inclusion
of the systematic errors would roughly double the total error.
We conclude that the analyses for the two subsets in EMF are in good
agreement, validating our choice to sum the two samples 
in the primary analysis.

\begin{figure}[!hbp]
\epsfig{figure=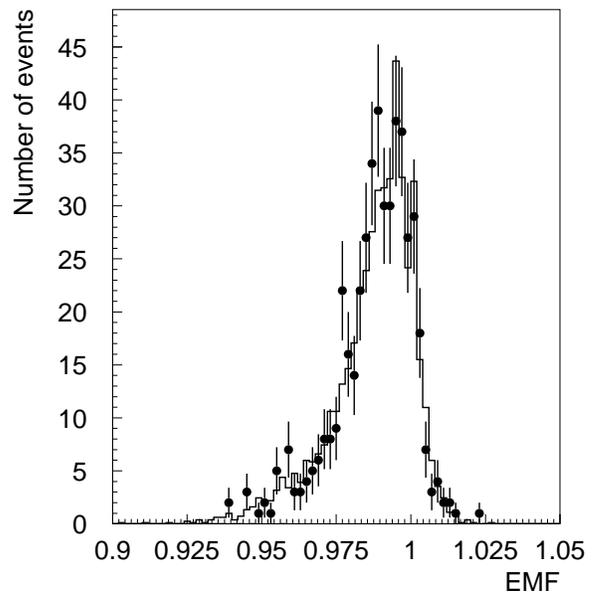,width=8.5cm}
\caption{
EM fraction distribution of edge electrons 
for the \edgec ~$W$ boson (data points)  and \edgecc ~$Z$ boson (histogram) 
samples.  The $Z$ boson sample is normalized to the $W$ boson sample.
}
\label{fig:emfoverlay}
\end{figure}

The averaging over the range of EMF values that occurs in our analysis
is acceptable if the electron EMF distribution is the same
for the \edgec ~$W$ boson sample and the $Z$ boson \edgecc ~sample
used to obtain the parameter values.  Fig.~\ref{fig:emfoverlay} shows the EMF
distributions for these two samples overlaid; they
are statistically consistent. 

\begin{table}
\caption{\label{tab:lohiemf}
Fitted parameters for edge electrons, and 
$W$ and $Z$ boson mass values, for
separate low and high EMF fraction subsamples.}
\begin{ruledtabular}
\begin{tabular}{c|cc} 
 ~ & Low EMF subset & High EMF subset \\ \hline
$\widetilde\alpha$    & ~~$0.922\pm 0.025$  & ~~$0.888\pm 0.024$ \\
$\widetilde c$       & ~~$0.163\pm 0.026$  & ~~$0.047\pm 0.027$ \\
$\widetilde f$       & ~~$0.45\pm 0.08$    & ~~$0.25\pm  0.06$  \\ \hline
$M_W$ (GeV) & $80.23\pm 0.34$ & $80.84\pm 0.29$ \\
$M_Z$ (GeV) & $91.43\pm 0.31$ & $90.96\pm 0.25$ \\
\end{tabular}
\end{ruledtabular}
\end{table}

\begin{figure}[!hbp]
\epsfig{figure=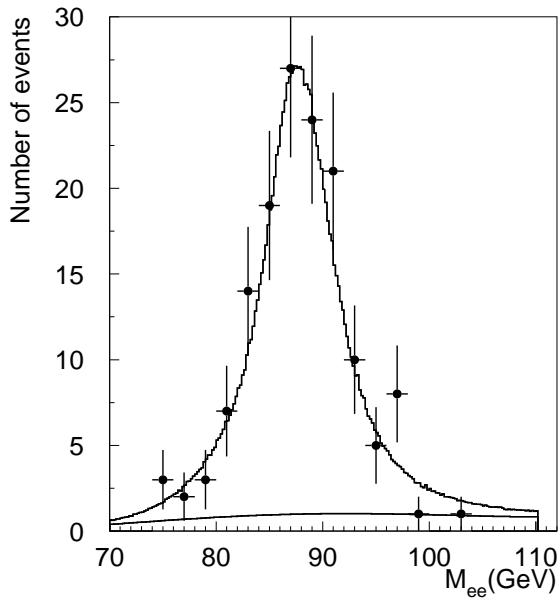,width=8.5cm}
\caption{
Best fit to the 
\edgece ~$Z$ boson mass distribution using
the parametrization discussed in the text for the central calorimeter
edge electron response 
and resolution,  and the parameterization of Ref. \cite{ecmw} for the end
calorimeter electron.  The histogram is the best fit from 
Monte Carlo, and the lower curve is the background.
}
\label{fig:cedgeendz}
\end{figure}

The parameters for edge electrons discussed above are determined
from the \edgecc ~$Z$ boson sample.  It is thus useful to
examine other samples in which \edgec ~electrons participate to
demonstrate the validity of the parametrization.   The \edgece
~dielectron sample with one edge central calorimeter electron and one
end calorimeter electron, using the energy response and resolution
of Ref. \cite{ecmw} for the end electrons, is shown in Fig.~\ref{fig:cedgeendz}.
This distribution is fit with $Z$ boson mass templates and
yields the result $M_Z=91.10 \pm 0.42$ GeV (statistical) with
$\chi^2=9.8$ for 13 degrees of freedom, in good agreement
with the precision LEP $Z$ boson mass determination.  When the reduced response
term for a fraction \fedge ~of central electrons in the edge region
is omitted, the fitted $Z$ boson mass is about one standard deviation
low, and the quality of the fit deteriorates to $\chi^2=11.7$.

\begin{figure}[!hbp]
\epsfig{figure=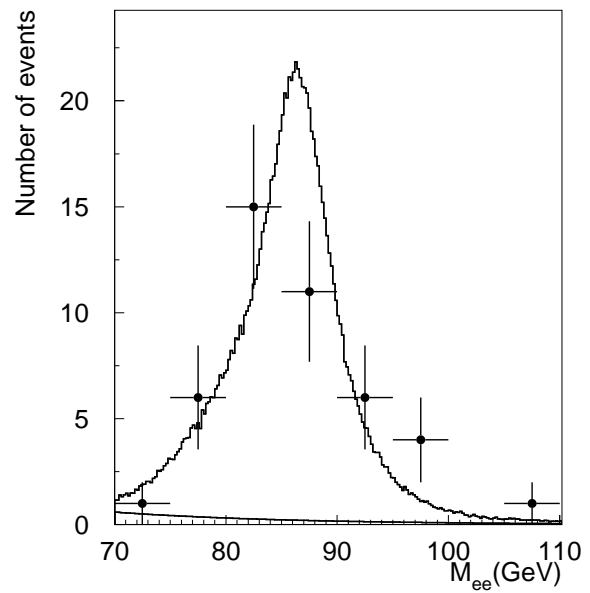,width=8.5cm}
\caption{
Best fit to the 
\edgeboth ~$Z$ boson mass distribution using
the parametrization discussed in the text for the central calorimeter
edge electron response and resolution.  The histogram is 
the best fit from 
Monte Carlo, and the lower curve is the background.
}
\label{fig:cedgecedgez}
\end{figure}

We also examine the dielectron sample in which both electrons
are in the central calorimeter edge region.  The data shown
in Fig.~\ref{fig:cedgecedgez} comprising 47 events is fitted to $Z$ boson 
mass templates
to give $M_Z=90.38 \pm 0.33$ GeV (statistical).   
The fit gives $\chi^2=8.5$ for 6 degrees of freedom.  When the
systematic errors are included, this result is in reasonable agreement
with the LEP precision value for $M_Z$.

\begin{figure}[!hbp]
\epsfig{figure=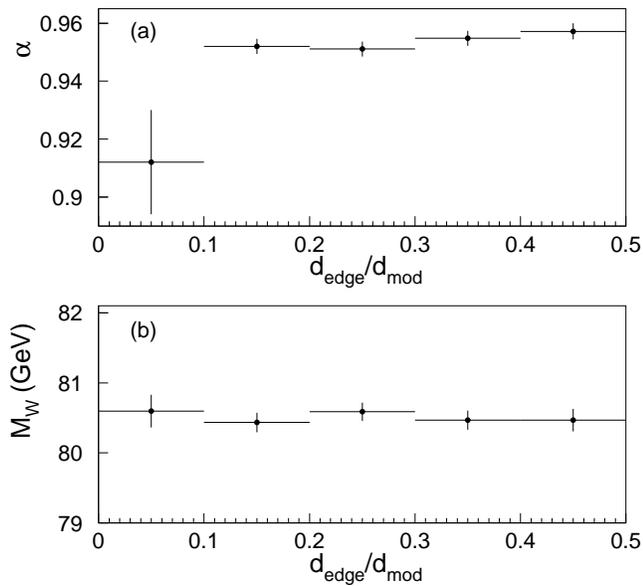,width=8.5cm}
\caption{
(a) The EM scale factor $\alpha$ and (b) the
fitted $W$ boson mass in bins of \dedge /$d_{\rm mod}$ using response
parameters from a $Z$ boson sample requiring one electron
in the same bin.
}
\label{fig:wmassdedge}
\end{figure}

As a final cross check, we subdivide the full $Z$
boson sample into five subsets, in which one electron (the `tagged'
electron) is required to be in a bin determined by the distance
\dedge ~from the nearest module edge.  
Five equal-sized bins span the range $0 < d_{\rm edge}/d_{\rm mod} < 0.5$.
The other electron is required to be in any of the non-edge
bins not populated by the tagged electron.   A companion sample
of $W$ boson candidates, subdivided into the five \dedge ~bins,
is also formed.    For each of the $Z$ boson samples, the tagged electron
response is fitted as described above with a variable energy
scale factor $\alpha$ using the LEP precision value as input.
This modified scale factor is then used for the $W$ boson subsamples
to obtain a best fit $W$ boson mass.  The results are shown
in Fig.~\ref{fig:wmassdedge}, where the points in the 
bin $0 < d_{\rm edge}/d_{\rm mod} < 0.1$ are 
those from the edge electron with additional parameters as described
above.  The resulting $W$ boson mass values are consistent 
over the five bins, indicating that our energy response correction
analysis is acceptable.

\section{Other parameter determinations}
\label{otherpar}

Although we expect that 
the main modifications to the previous non-edge electron
$W$ boson analyses are the response and 
resolution parametrizations discussed in Section \ref{response}, there are some
other parameters that could be sensitive to the location of the
electron relative to the module boundary.

The observed electron and recoil system energies
are changed from the true values by the energy 
from the underlying event deposited in the region
used to define the electron.  This component
of energy must be subtracted from the observed electron energy and
added to the recoil.  In Ref. \cite{ccmw1b} we found this correction
to be dependent on the electron rapidity and on the instantaneous
luminosity.   The size of the region used to collect the electron
energy is $\Delta\eta \times \Delta\phi = 0.5\times0.5$, spanning
two and a half times the size of a module in the $\phi$ direction.  
Thus the underlying event correction can only be very weakly dependent on the
location of the center of this region, and we take the correction to
be the same as for the non-edge analysis.  Also, the recoil
system has its momentum vector pointing anywhere in the detector in
both the edge and non-edge analyses.  Thus we do not modify the previous
parameters controlling the recoil system response and resolution.

The efficiency for finding electrons is modified by the underlying
event energy within the electron region.
The efficiency depends on $u_\parallel$, since when there is
substantial recoil energy near the electron, the isolation requirement
will exclude more events than when the recoil energy
is directed away from the
electron.   Since the electron energy itself is modified
near the module edge, this efficiency could be different for C
and \edgec ~electrons.   To investigate this effect, we compute
the average $f_{\rm iso}$ for both C and \edgec ~samples.  We find
that $\langle f_{\rm iso} \rangle$ for the \edgec ~ sample is
$1.08 \pm0.15$ times that for the C sample.   We expect about
a 3\% increase in $\langle f_{\rm iso} \rangle$ since its definition
involves the EM energy near the core of the shower, 
which is reduced for \edgec ~electrons.    A modified distribution
of $f_{\rm iso}$ can only affect the $u_\parallel$ efficiency 
if there is a change in the $u_\parallel$ distribution in the 
\edgec ~events relative to that for the C electrons.  
We see no difference in the $\langle f_{\rm iso} \rangle$
value in hemispheres $u_\parallel < 0$ and $u_\parallel > 0$ for the 
\edgec ~events.  This
observation, and the statistically insignificant difference
for $\langle f_{\rm iso} \rangle$ for C and \edgec ~samples, lead us to
retain the previous parametrization for the $u_\parallel$ efficiency.

Since photons radiated from electrons are found
dominantly near the electron, these photons also populate
reduced response regions in the edge electron analysis.
For our analysis we have chosen to generate such
radiation with the response parameters found for the \edgec ~electrons.
However some of the radiated $\gamma$'s strike the non-edge region and
should thus be corrected with the non-edge response.  We calculate
that the difference between the photon
energy using the edge response 
and a properly weighted response across the module is only
3.5 MeV, resulting in a negligible less  shift in the $W$ boson
mass \cite{slava}.    

When an electron impacts the calorimeter near a cell boundary, as
occurs near the module edge, its position resolution in $r\phi$ is
improved typically by about 20\% \cite{d0nim}.  
This means that the determination of the electron cluster azimuth 
is more accurate for \edgec ~than for C electrons.
The effect of improved azimuthal precision in the \edgec ~sample has
however been incorporated by fitting the energy response and
resolution parameters for the \edgecc ~$Z$ boson sample, so no
additional correction is needed.

The small modification to the electron energy (a 4\% reduction in 35\%
of the electrons in the edge region) could affect the
trigger efficiency near the threshold.   We determine that this
effect is negligible.

\section{$W$ boson mass determination}
\label{wmass}

\subsection{Mass fits}

\begin{figure}[!hbp]
\epsfig{figure=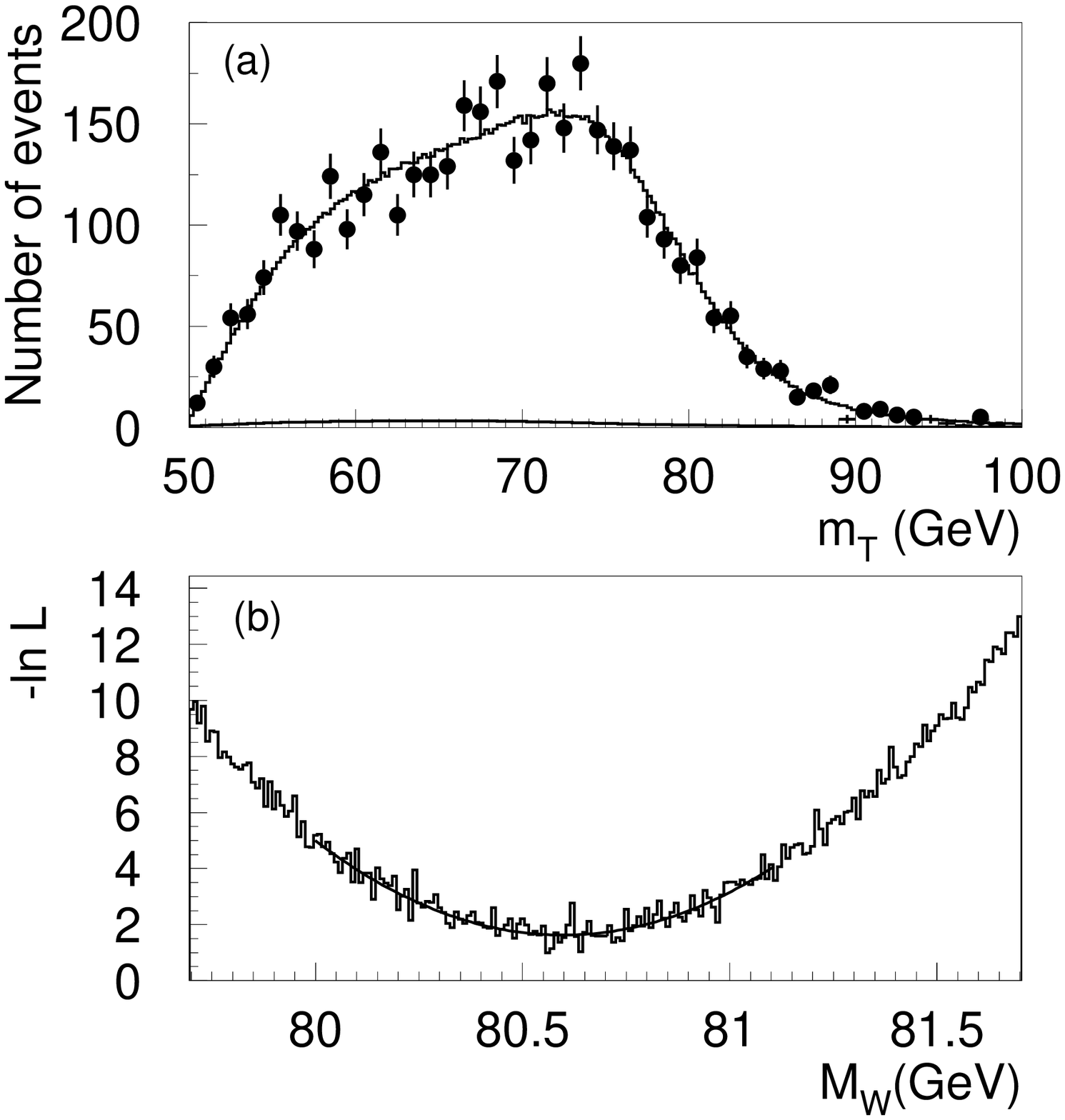,width=8.5cm}
\caption{
(a) Comparison of the data (points)
and the Monte Carlo predicted distribution (histogram)
in transverse mass,
using the fitted value for $M_W$.  The Monte Carlo distribution is
normalized in area to the number of $W$ boson events within
the fitting window.  The 
estimated backgrounds are indicated by the lower curve.
(b) The distribution of calculated likelihood values as
a function of the assumed $W$ boson mass.  The curve is
a fitted parabola.
}
\label{fig:mtwfit}
\end{figure}

\begin{figure}[!hbp]
\epsfig{figure=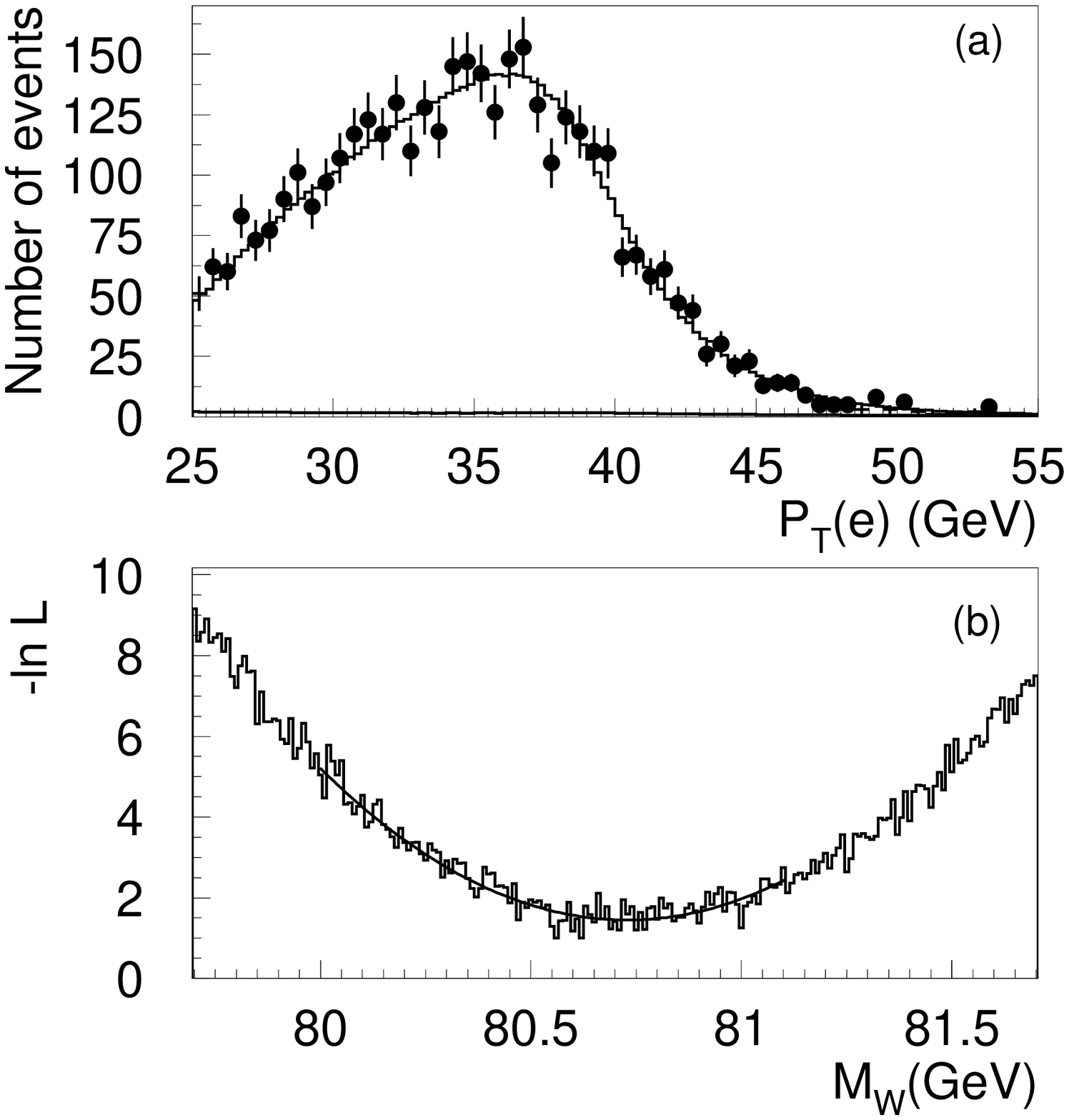,width=8.5cm}
\caption{
(a) Comparison of the data (points)
and the Monte Carlo predicted distribution (histogram)
in electron transverse momentum,
using the fitted value for $M_W$.  The Monte Carlo distribution is
normalized in area to the number of $W$ boson events within
the fitting window.  The 
estimated backgrounds are indicated by the lower curve.
(b) The distribution of calculated likelihood values as
a function of the assumed $W$ boson mass. The curve is
a fitted parabola.
}
\label{fig:ptefit}
\end{figure}

\begin{figure}[!hbp]
\epsfig{figure=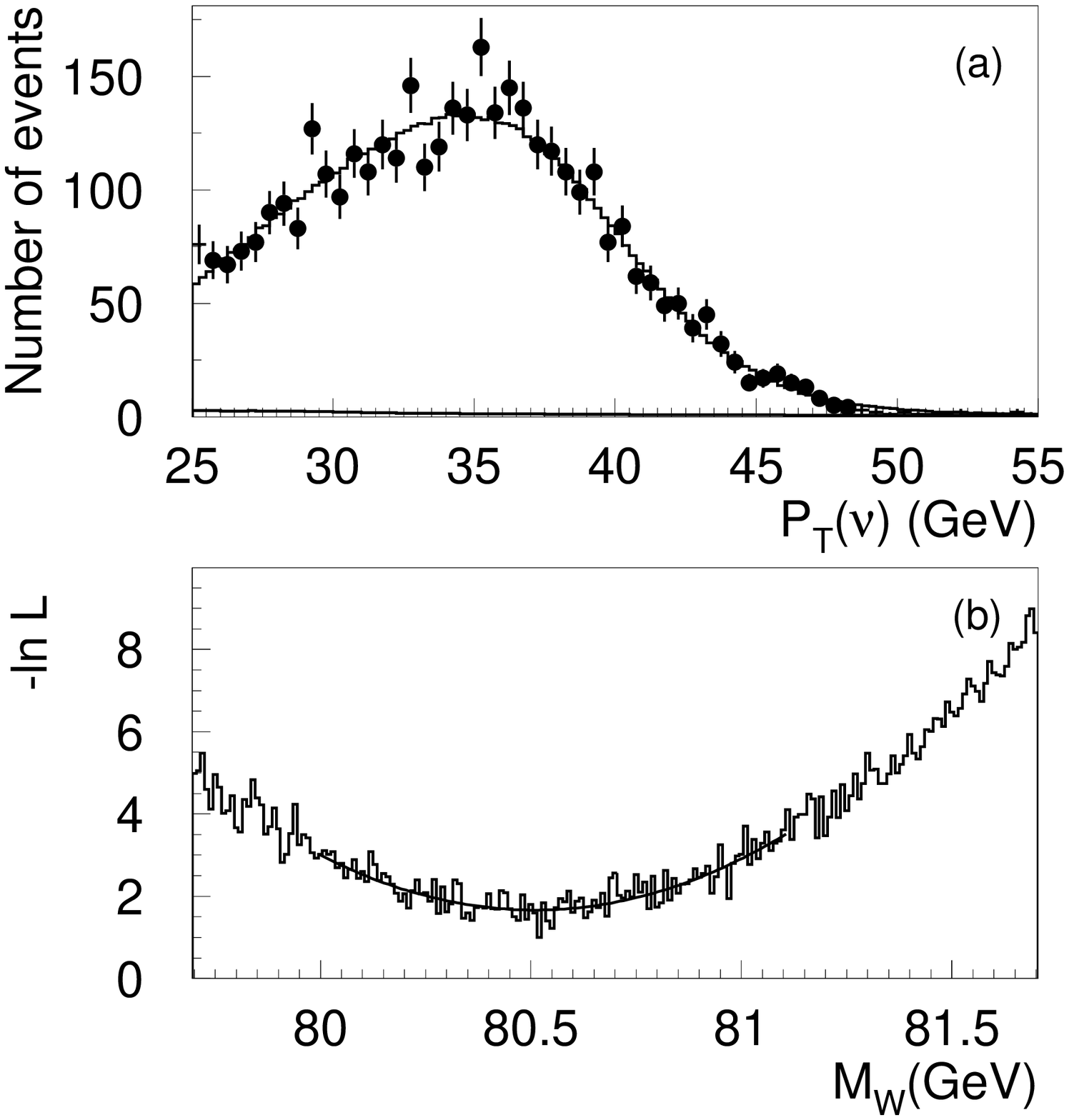,width=8.5cm}
\caption{
(a) Comparison of the data (points)
and the Monte Carlo predicted distribution (histogram)
in neutrino transverse momentum,
using the fitted value for $M_W$.  The Monte Carlo distribution is
normalized in area to the number of $W$ boson events within
the fitting window.  The 
estimated backgrounds are indicated by the lower curve.
(b) The distribution of calculated likelihood values as
a function of the assumed $W$ boson mass. The curve is
a fitted parabola.
}
\label{fig:ptnufit}
\end{figure}

Monte Carlo templates are prepared for the $W$ boson
transverse mass $m_T$, electron
transverse momentum $p_T(e)$, and neutrino transverse
momentum $p_T(\nu)$, using the production, decay, and detector
parameters discussed in Sections \ref{experiment} and \ref{response}.  
The estimated backgrounds
described in Section \ref{background} 
are added to the Monte Carlo $W$ boson decays.
Families of templates are
made for $W$ boson masses varied in 10 MeV steps between 79.6 and 81.6
GeV.  The templates are compared to the data in the ranges 
$60 \leq m_T < 90$ GeV, $30 \leq p_T(e) < 50$ GeV, and
$30 \leq p_T(\nu) < 50$ GeV, with bins of 100 MeV for
transverse mass and 50 MeV for the transverse momentum
distributions.
For each specific template with fixed $M_W$, we normalize the 
distributions to the data 
within the fit interval and compute a binned likelihood
\begin{equation}
 \mathcal{L}(m) = \prod_{i=1}^N p_i^{n_i}(m)
\end{equation}
where $p_i(m)$ is the probability density for bin $i$ with the $W$
boson mass taken as $m$,
$n_i$ is the number of data events in bin $i$, and $N$ is the number
of bins in the fit interval. We fit $-\ln \mathcal{L}(m)$ 
with a quadratic function of
$m$. The value of $m$ at which the function assumes its minimum is the fitted
value of the $W$ boson mass and the 68\% confidence level statistical
error corresponds to the
interval in $m$ for which $-\ln \mathcal{L}(m)$ 
is within half a unit of the minimum.
The best fit $m_T$, $p_T(e)$ and $p_T(\nu)$
distributions and the associated likelihood curves 
are shown in Figs.~\ref{fig:mtwfit} -- \ref{fig:ptnufit}.
The fitted values for $M_W$ and $\chi^2$ from each of the distributions
are given in Table~\ref{tab:fitmass}.  
The errors shown are statistical only; the
values of $M_W$ obtained from the three distributions are in good
agreement.

\begin{figure}[!hbp]
\epsfig{figure=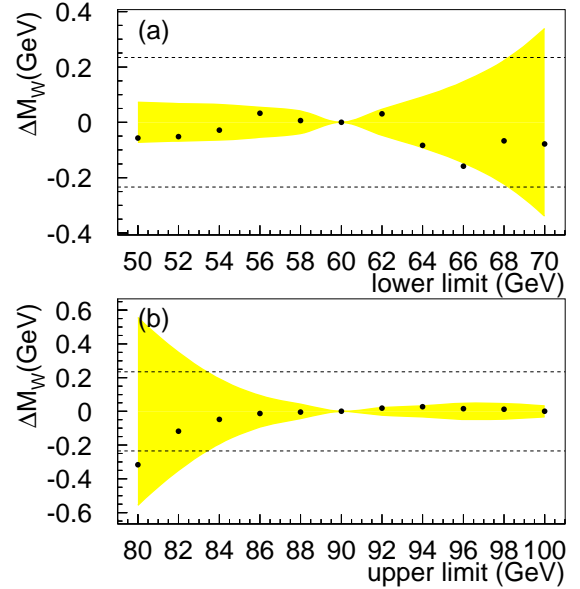,width=8.5cm}
\caption{
Variation of the fitted $W$ boson mass with (a) the lower edge 
and (b) the upper edge of the fit window for the transverse
mass distribution.   The shaded regions and the dashed lines
are described in the text.
}
\label{fig:mtwindow}
\end{figure}

We study the sensitivity of the fits to the choice of fitting
window by varying the upper and lower window edges by $\pm 10$
GeV for the transverse mass and by $\pm 5$
GeV for the transverse momentum fits.  Figure~\ref{fig:mtwindow}
shows the change in $M_W$ as the upper and lower window edges for
the transverse mass fit are varied.  The shaded bands correspond 
to the 68\% probability contours, determined from an ensemble
of Monte Carlo $W$ boson samples with the chosen window edges.
The dashed lines indicate the statistical error for the nominal fit.
The points for different window edges are correlated, as the data
with a larger window contains all the data in a smaller window.
The deviations of $M_W$ are in good agreement for differing choices of
window.  Similar good agreement is seen in varying the windows
for the $p_T(e)$ and $p_T(\nu)$ fits.

\begin{table}
\caption{\label{tab:fitmass}
Fitted $W$ boson masses and $\chi^2$/degrees of freedom.}
\begin{ruledtabular}
\begin{tabular}{c|cc} 
 Distribution & Fitted mass & $\chi^2$/d.o.f. \\ \hline
 $m_T$ & $80.596\pm 0.234$   & 45/29 \\
 $p_T(e)$ & $80.733\pm 0.263$   & 38/39 \\
 $p_T(\nu)$ & $80.511\pm 0.311$   & 45/39 \\
\end{tabular}
\end{ruledtabular}
\end{table}

\subsection{Mass error determination}

In addition to the statistical errors determined from the fits,
there are systematic errors arising from the uncertainties in 
all of the parameters that enter in the Monte Carlo production, decay
and detector model.   These parameters, summarized in 
Table~\ref{tab:parameters}, form a parameter
vector $\vec P$.   The definition and determination of the parameters
are described above and in Ref. \cite{ccmw1b}.
The recoil response takes into account the joint 
effects of two correlated parameters $\alpha_{\rm rec}$ and
$\beta_{\rm rec}$.  We assign an uncertainty in $M_W$ for
the uncorrelated errors obtained from the principal axes of the 
$\alpha_{\rm rec} - \beta_{\rm rec}$ error ellipse \cite{ccmw1b}.
The recoil resolution depends on correlated parameters
$s_{\rm rec}$ and $\alpha_{\rm mb}$ \cite{ccmw1b}, and the
$u_\parallel$ efficiency depends on correlated parameters
$u_0$ and $s_0$; these correlated pairs are treated similarly to
those for the recoil response.
The set of production model errors include the parameters
due to parton distribution function (PDF) uncertainty, $W$ boson 
width \cite{wwidth},
the parameters determining the $W$ boson production $p_T$ spectrum, and
the parton luminosity function.   We take the components of the 
production model error to be uncorrelated.
The PDF error is taken from the deviation of the $W$ boson mass 
comparing \cite{ecmw} 
MRS(A)$^\prime$ \cite{MRSA}, 
MRSR2 \cite{MRSR2}, 
CTEQ5M \cite{CTEQ5M}, 
CTEQ4M \cite{CTEQ4M}, 
and CTEQ3M \cite{CTEQ3M} PDF's
to our standard choice of MRST.

In all, we have identified $N_P=21$ 
parameters that determine the model for the
Monte Carlo:  the eighteen used in the previous studies and the
three new parameters related to the edge electrons (\alphaedge ,
\cedge ~and \fedge ).

\begin{table}
\caption{\label{tab:parameters}
Parameters $\vec P$  used in the $W$ boson mass determination}
\begin{ruledtabular}
\begin{tabular}{|c|c|} 
 Parameter & Description\\ \hline
 $\alpha$ & EM energy response scale for non-edge $e$ \\
 $\widetilde \alpha$ & EM energy response scale for edge $e$\\
 $\delta$  & EM response offset \\
 $c$ & EM resolution constant for non-edge $e$ \\
 $\widetilde c$ & EM resolution constant for edge $e$ \\
 $\widetilde f$ & fraction of low response $e$ in edge region   \\
 $\beta_{\rm cdc}$ & drift chamber position scale factor  \\
 $\alpha_{\rm rec}$ & recoil energy response scale constant  \\
 $\beta_{\rm rec}$ & recoil energy response scale $Q^2$ dependence  \\
 $s_{\rm rec}$ & recoil energy resolution  \\
 $\alpha_{\rm mb}$    & recoil energy from added minimum bias events  \\
 $\Delta u_\parallel$ & underlying event energy correction in $e$ window  \\
 $u_0$ & $u_\parallel$ cutoff for constant efficiency\\
 $s_0$ & slope of $u_\parallel$ efficiency vs $u_\parallel$ \\
 $b_W$ & background to $W$ boson distribution  \\
 $r_\gamma$ & coalescing radius for photon radiation  \\
 $2\gamma$ & error for 2 $\gamma$ radiation  \\
 PDF  & error from varying PDF   \\
 $\Gamma_W$ & $W$ boson width  \\
 $\beta$ & parton luminosity   \\
 $g_2$ & $Q^2$ dependence of $W$ boson production  \\
\end{tabular}
\end{ruledtabular}
\end{table}

The parameters ${P_i}$ are determined from 
$N_Y = 32$ auxiliary measurements
using several data sets 
which include the CC and \edgecc ~$Z$ boson samples, 
special minimum bias and muon samples for determining drift
chamber scales and underlying event properties, and external
data sets that are used to constrain the $W$ boson production model.
The measurements using these special data sets are
denoted ${Y_I}$ ($I=1,...~N_Y$) with
uncertainties ${\sigma^Y_I}$. 
Each measurement puts constraints on one or more of the 
parameters $P_i$.  Measurements ${Y_I}$ are related to
the parameters ${P_i}$ through the functional relation
$Y_I = F_I(\vec{P})$.

We form the $\chi^2$ for the set of measurements
\begin{equation}
\chi^2 = \sum_{I,J=1}^{N_Y}
{[Y_I - F_I(\vec{P})](C^Y_{IJ})^{-1}[Y_J - F_J(\vec{P})]
\over (\sigma^Y_I\sigma^Y_J)} ~,
\label{eq:chisqY}
\end{equation}
\noindent where $C^Y_{IJ}
=\langle\Delta Y_I \Delta Y_J\rangle$ 
is the covariance matrix of the measurements, determined from 
Monte Carlo calculations.
If the deviations of the measurements from their means 
are taken to be linearly related
to the parameters in the region of the $\chi^2$ minimum:
\begin{equation}
\Delta Y_I = \sum_{j=1}^{N_P}{D^Y_{Ij}\Delta P_j} ,
\label{eq:deriv}
\end{equation}
\noindent 
where $D^Y_{Ij}=\partial F_I/\partial P_j$, 
the minimum of the $\chi^2$ can be found analytically.
The parameter covariance matrix ${C^P_{ij}}$
can be then calculated from 
$C^Y_{IJ}$ and the 
derivatives $D^Y_{Ij}$.

This analysis is carried out for the three distinct measurements
of $M_W$ for the edge electrons
($m_T$, $p_T(e)$ and $p_T(\nu)$).  Each
measurement depends on the set of parameters, $\vec P$,
discussed above.
For the $N_M=3$ 
separate mass measurements ${m_\alpha}$ ($\alpha=1,...N_M$), the
mass measurement covariance matrix $C^M_{\alpha\beta}$ is obtained from 
\begin{equation}
C^M_{\alpha\beta}=\sum_{k,l=1}^{N_P}D^M_{\alpha k} C^P_{kl} D^M_{l\beta} ,
\end{equation}
where $D^M_{\alpha j} = \partial m_\alpha/\partial P_j$.   
The correlation of the statistical errors is obtained
from studies of Monte Carlo ensembles; these correlations
are shown in Table~\ref{tab:statcorr}.

We can fit for the best
combined mass value $M_W$ by minimizing the $\chi^2$ \cite{lyons}
\begin{equation}
\chi^2=\sum_{\alpha,\beta=1}^{N_M} (m_\alpha - M_W) H_{\alpha\beta} 
    (m_\beta-M_W) ,
\end{equation}
where $\mathbf{H} = (\mathbf{C^M})^{-1}$.
The best fit is given by
\begin{equation}
M_W=\bigl(\sum_{\alpha,\beta=1}^{N_M} H_{\alpha\beta}m_i\bigr)/
    \sum_{\alpha,\beta=1}^{N_M} H_{\alpha\beta} ,
\end{equation}
with error 
\begin{equation}
\sigma_m=(\sum_{\alpha,\beta=1}^{N_M} H_{\alpha\beta})^{-1/2} .
\end{equation}

\begin{table}
\caption{\label{tab:statcorr}
The statistical correlation coefficients for the
three measurements of the $W$ boson mass.
}
\begin{ruledtabular}
\begin{tabular}{cccc} 
~ & $m_T$ & $p_T(e)$ & $p_T(\nu)$ \\ \hline
$m_T$    &  1 & 0.669   & 0.630 \\
$p_T(e)$ & 0.669 & 1 & 0.180 \\
$p_T(\nu)$ & 0.630 & 0.180 & 1 \\
\end{tabular}
\end{ruledtabular}
\end{table}

The resultant $W$ boson mass measurements using electrons in the
edge region are
\begin{equation}
M_W=80.596\pm 0.234\pm 0.370 ~{\rm GeV}
\end{equation}
\noindent for the $m_T(W)$ fit,
\begin{equation}
M_W=80.733\pm 0.263 \pm 0.460 ~{\rm GeV}
\end{equation}
\noindent  for the $p_T(e)$ fit, and
\begin{equation}
M_W=80.511\pm 0.311 \pm 0.523 ~{\rm GeV}
\end{equation}
for the $p_T(\nu)$ fit, where the first error is statistical and the second is
systematic.
\noindent
The breakdown of the contributions to the systematic errors
is shown in Table~\ref{tab:errors}.
The PDF error is taken as the difference on the combined $W$ boson
mass between the CTEQ3M and MRST choices, for which $m_W$ differs
maximally.  The combined mass error from this source (not shown
in Table~\ref{tab:errors}) is 19 MeV.
The errors associated with the broad Gaussian
parameters  in the
edge electron response (\alphaedge ~and \cedge ) 
dominate the systematic errors.

The three measurements of $M_W$ are correlated as shown
in Table~\ref{tab:mwcorr}; when combined taking these correlations
into account, we obtain
\begin{equation}
M_W=80.574\pm 0.405 ~{\rm GeV} ,
\end{equation}
\noindent
with $\chi^2=0.61$ for two degrees of freedom.

\begin{table}
\caption{\label{tab:errors} Errors (in MeV) for the three 
$W$ boson measurements}
\begin{ruledtabular}
\begin{tabular}{|c|r|r|r|}
 Source          &  \mt &  \pte  & \ptnu \\ \hline
 statistics                       &  234  &  263   & 311 \\ 
 edge EM scale ($\widetilde \alpha$) &  265  &  309   & 346 \\ 
 CC EM scale  ($\alpha$)     &  128  &  131   & 113 \\ 
 CC EM offset ($\delta$)     &  142  &  139   & 145 \\ 
 calorimeter uniformity           &  10   &  10    & 10  \\ 
 CDC scale                        &  38   &  40    & 52  \\ 
 backgrounds                      &  10   &  20    & 20  \\ 
 CC EM constant term $c$     &  15   &  18    & 2   \\ 
 edge EM constant term ($\widetilde c$)   &  268  &  344   & 404 \\ 
 fraction of events  ($\widetilde f$)     &   8   &  14    & 22  \\ 
 hadronic response                &  20   &  16    & 46  \\ 
 hadronic resolution              &  25   &  10    & 90  \\ 
 \upar~ correction                &  15   &  15    & 20  \\ 
 \upar~ efficiency                &  2    &  9     & 20  \\ 
 parton luminosity                &  9    &  11    & 9   \\ 
 radiative corrections            &  3    &  6     & $<1$   \\ 
 $2\gamma$                        &  3    &  6     & $<1$   \\ 
 $p_T(W)$ spectrum                & 10    &  50    & 25  \\ 
 $W$ boson width                  & 10    &  10    & 10  \\ 
\end{tabular}
\end{ruledtabular}
\end{table}

\begin{table}
\caption{\label{tab:mwcorr}
The full correlation coefficients for the
three measurements of the $W$ boson mass.
}
\begin{ruledtabular}
\begin{tabular}{cccc} 
~ & $m_T$ & $p_T(e)$ & $p_T(\nu)$ \\ \hline
$m_T$    &  1 & 0.90   & 0.89 \\
$p_T(e)$ & 0.90 & 1 & 0.76 \\
$p_T(\nu)$ & 0.89 & 0.76 & 1 \\
\end{tabular}
\end{ruledtabular}
\end{table}

\section{Combination of all \d0 $W$ boson mass measurements}
\label{combination}
The analysis presented here for the edge electrons brings two
new ingredients to the \d0 $W$ mass measurements.   First,
the edge electron sample is statistically independent of all other
measurements, and thus can be combined to give an improved $M_W$
measurement.  Second, the added statistics of the \edgecc ~and
\edgece ~$Z$ boson samples can be used to refine the knowledge 
of the electron response parameters for \textit{non-edge}
central calorimeter or end calorimeter electrons.
The improved energy scale factors in turn give improved $W$ 
boson mass precision.

\subsection{Modified non-edge electron $W$ boson mass}
Using the \edgecc ~sample and the same fitting procedure described
in Section~\ref{response} for the \edgec ~electrons, we have obtained a
scale factor $\alpha = 0.9552\pm0.0023$ for the \textit{non-edge}
electrons.  This value
can be compared with the previous 
determination from the CC sample \cite{ccmw1b}
of $\alpha=0.9540\pm0.0008$.
The correlation matrix for CC and \edgecc ~measurements 
is calculated in the manner discussed in Section VI.

Similarly, the \edgece ~sample can be used to constrain the scale
factor $\alpha$ for both end and non-edge central electrons (recall that the
central edge electrons contain a fraction $(1-f_{\rm edge})$ of events
whose scale factor and resolution are identical to those of the
central non-edge electrons).  Taking into account the correlations, 
we obtain $\alpha=0.9559\pm  0.0107$ for electrons in the non-edge
region of the central calorimeter and $\alpha=0.9539\pm0.0085$ for
the electrons in the end calorimeter.  The latter value can
be compared with the previous value \cite{ecmw} of the end calorimeter
electron scale of $0.9518\pm0.0019$.

Taking the two new measurements of $\alpha$ for the central
calorimeter together with the previously determined value, we obtain 
\begin{equation}
\alpha=0.9541\pm 0.00075  .
\end{equation}
This new scale factor
is higher than the previous value by $0.0001$, and the
error is reduced by 6\%.  For the end calorimeter, the new
combined scale factor is
\begin{equation}
\alpha=0.9519\pm 0.0018  ,
\end{equation}
again higher than the previous value by $0.0001$ with a 5\% reduction
in error.  

In principle, the added data could also improve the
precision for the resolution constant term $c$ in 
the central and end calorimeters, but in practice it does not.

With the new values for the scale factors for the non-edge central
calorimeter electrons, we obtain
modified results for the non-edge central calorimeter $W$ boson
mass:
\begin{equation}
M_W=80.438 \pm 0.107~{\rm GeV} ,
\end{equation}
to be compared with the published value of
$M_W=80.446 \pm 0.108$ GeV \cite{ccmw1b}.
The new end calorimeter electron scale factor gives a modified $W$ boson mass:
\begin{equation}
M_W=80.679 \pm 0.209~{\rm GeV} ,
\end{equation}
to be compared with the published value from the
end calorimeters of
$M_W=80.691 \pm 0.227$ GeV \cite{ecmw}.

With the modified scale factors for C and E electrons, we obtain
\begin{equation}
M_W=80.481\pm 0.085 ~{\rm GeV} ,
\end{equation}
with $\chi^2=5.5$ 
(6 degrees of freedom) for all non-edge central 
and end calorimeter measurements,
compared with the previous determination  
$M_W=80.482\pm 0.091$ GeV \cite{ecmw}.

\subsection{Combined $W$ boson mass from all \d0 measurements}
With the edge electron mass determinations reported in this paper,
there are now ten separate \d0 $W$ boson measurements: the Run 1a
central calorimeter transverse mass measurement \cite{ccmw1a}, three
Run 1b 
central calorimeter non-edge measurements \cite{ccmw1b} (from the transverse
mass and electron and neutrino transverse momenta), three Run 1b
end calorimeter measurements \cite{ecmw}, and the three present
measurements of the central calorimeter edge electrons.
Combining these ten mass measurements using the method outlined
in Sec.~\ref{wmass} and an expanded set of measurements and parameters 
to incorporate also the
end calorimeter electrons, we obtain a 
final \d0 combined measured value for the $W$ boson mass of
\begin{equation}
M_W=80.483 \pm 0.084 ~{\rm GeV}
\end{equation}
with $\chi^2=6.3$ (9 degrees of freedom).
This value is to be compared with our previous \cite{ecmw} combined measurement
of $M_W=80.482\pm0.091$ GeV.   The edge electrons in the central
calorimeter have improved the precision over the previously published
results by 7 MeV, or 8\%.

\section{Summary}
\label{summary}

Using a sample of electrons which impact upon 
the 10\% of a central calorimeter module closest to either module edge
in azimuth, we have made a new measurement of the $W$ boson mass,
and have refined our knowledge of the energy scale for
previously used electrons that are in the interior 80\% of the central
calorimeter modules or are in the end calorimeters.   
Adding the new measurement using the edge electrons gives the
final combined result
$$M_W=80.483\pm0.084 ~{\rm GeV} ~~~({\rm D\O}).$$

Combining the new \d0 $W$ boson mass value reported here with
the CDF \cite{cdfmw} and UA2 \cite{ua2} measurements, 
taking into account the updated correlated systematic errors
for the three experiments due to
parton distribution function uncertainties and multiple photon
radiation gives \cite{tevavg}
$$M_W=80.454\pm0.059 ~{\rm GeV} ~~~(p\overline p).$$
This is an improvement over the previous measurement from hadron
colliders of $M_W=80.452 \pm 0.062$ GeV \cite{kotwal}.
Further combining with the LEP experiments' 
preliminary measurement $M_W=80.450\pm 0.039$ GeV \cite{ewwkgp},
we find the world average $W$ boson mass from direct 
measurements to be \cite{tevavg}
$$M_W=80.451\pm0.033 ~{\rm GeV} ~~~({\rm world}).$$

The edge electrons used in this analysis represent a 
14\% increase in the central
calorimeter $W$ boson sample, and an 18\% increase in the total
$Z$ boson sample.   The larger sample sizes should be of use
for all subsequent studies of vector bosons in \d0.

\section*{ Acknowledgements }

%
We thank the staffs at Fermilab and collaborating institutions, 
and acknowledge support from the 
Department of Energy and National Science Foundation (USA),  
Commissariat  \` a L'Energie Atomique and 
CNRS/Institut National de Physique Nucl\'eaire et 
de Physique des Particules (France), 
Ministry for Science and Technology and Ministry for Atomic 
   Energy (Russia),
CAPES and CNPq (Brazil),
Departments of Atomic Energy and Science and Education (India),
Colciencias (Colombia),
CONACyT (Mexico),
Ministry of Education and KOSEF (Korea),
CONICET and UBACyT (Argentina),
The Foundation for Fundamental Research on Matter (The Netherlands),
PPARC (United Kingdom),
Ministry of Education (Czech Republic),
A.P.~Sloan Foundation,
NATO, and the Research Corporation.

\end{document}